\documentclass[preprint,11pt]{elsarticle}
\usepackage[utf8]{inputenc}
\usepackage[T1]{fontenc}
\usepackage{lmodern}
\usepackage{amsmath,amssymb,bm}
\usepackage{graphicx}
\usepackage{booktabs}
\usepackage{subcaption}
\usepackage[margin=2.5cm]{geometry}
\usepackage[hidelinks]{hyperref}
\newcommand{\V}{\mathbf V}
\newcommand{\x}{\mathbf x}
\newcommand{\n}{\mathbf n}
\newcommand{\q}{\mathbf q}

\makeatletter
\def\ps@pprintTitle{\let\@oddhead\@empty\let\@evenhead\@empty
  \def\@oddfoot{\footnotesize\itshape Technical report, CNR-IAC\hfill\today}%
  \let\@evenfoot\@oddfoot}
\makeatother

\begin{document}
\begin{frontmatter}

\title{BBun: an unsteady vortex-ring panel method for interacting lifting
bodies, box wings and propellers\\[4pt]
\large Formulation, implementation, verification and validation\\
Technical report}

\author{Daniele Peri}
\cortext[cor]{Corresponding author: d.peri@iac.cnr.it}
\address[inst]{\small CNR-IAC -- National Research Council \\
      Istituto per le Applicazioni del Calcolo "Mauro Picone" \\
      Via dei Taurini 19, 00185 Rome, Italy}

\begin{abstract}
This report documents BBun (Blunt Body unsteady), a low-order code for the
unsteady potential flow around one or more interacting three-dimensional
bodies. The
bodies are represented by vortex rings (piecewise-constant doublets) with a
Neumann boundary condition; free vortex sheets are shed from the trailing
edges and, optionally, from the tips, and convected in time, with an
implicit unsteady Kutta condition. The formulation is classical and
deliberately simple. The report describes the choices that make the
discretization consistent and convergent (a single desingularized kernel per
wake line, a bordered factorization of the singular Neumann problem,
interior velocity conditions for the nearly singular mode of ring-shaped
bodies, the detection of the separation lines from the grid connectivity)
and the extensions to several bodies, box wings, symmetric configurations
(image method), non-lifting bodies, rigidly rotating bodies (propellers), the
unsteady Bernoulli equation, a wake--body collision treatment and a
strip-wise integral boundary layer. The verification and validation are
reported in full: potential flow around a sphere, sensitivity and grid
convergence, the two-dimensional limit, an independent vortex-lattice
method, the NACA~0015 wing of McAlister and Takahashi (loads and chordwise
pressures), the NACA~0012 profile drag, the biplane of Knight and Noyes,
Prandtl's best wing system, Wagner's function and blade-element momentum
theory for a propeller. The limitations of the model are stated
explicitly; the most important is that, with this discretization, the
surface velocity and hence the pressure lift have a first-order error in the
chordwise panel size (about $-130/n$\,\% with $n$ panels per side), while
the circulation converges fast, so that lift-to-drag ratios, span
efficiencies and propeller loads are computed from the circulation. A
complete aircraft with counter-rotating tractor propellers illustrates the
use of the code.
\end{abstract}

\end{frontmatter}

\section{Introduction}

Low-order potential-flow methods remain a standard tool in the conceptual and
preliminary design of lifting systems, where many configurations must be
evaluated at a small fraction of the cost of a viscous simulation. Surface
panel methods, from the source method of Hess and Smith~\cite{hess1967} to
the doublet formulations of Morino and Kuo~\cite{morino1974} and the codes
VSAERO~\cite{maskew1987}, PMARC~\cite{ashby1991} and, with higher-order
panels, PAN~AIR~\cite{magnus1980}, represent the thickness of the body
within potential theory; see Refs.~\cite{hess1990,erickson1990} for reviews.
Unsteady vortex-lattice and panel methods with a free wake, as presented by
Katz and Plotkin~\cite{katz2001}, add the time history of the shed
vorticity and a force-free wake geometry. Current mid-fidelity tools for
complete aircraft with propellers combine such surface models with lifting
lines, vortex lattices and vortex particles (e.g.\ DUST~\cite{tugnoli2021},
VSPAERO within OpenVSP~\cite{mcdonald2022}), and vortex particle methods
resolve the wake with viscous diffusion and stretching~\cite{winckelmans1993,alvarez2020}.

The method documented here belongs to the simplest of these classes: a
first-order, constant-strength vortex-ring discretization with the Neumann
condition, a lattice of vortex rings as free wake, convected with an
explicit scheme, and a strip-wise integral boundary layer. It is not a
competitor of the tools above, and its formulation is not new; its interest
lies in being simple, fully documented and verified, and adapted to the
configurations for which it has been developed: closed lifting systems such
as the box wing, several interacting bodies, and propellers in front of
lifting surfaces. The box wing, in which a lower and an upper wing are
joined at their tips, is the lifting system of minimum induced drag for
given span and lift, as conjectured by Prandtl~\cite{prandtl1924} and shown
by Frediani and Montanari~\cite{frediani2009}; it has no tips and its wake is
a closed band, which requires the identification of a closed trailing edge
and a sheet without free edges.

This report is the reference for the code BBun and for the applications that
use it. It describes the model (Section~\ref{sec:model}) and the numerical
method (Section~\ref{sec:method}), states the scope and the limitations of
the model (Section~\ref{sec:scope}), reports the verification and validation
in full (Section~\ref{sec:results}), and gives the recommendations for its use
that follow from them (Section~\ref{sec:concl}). The discretization choices
that required care, and that are the specific content of this work, are:
\begin{itemize}
\item an implicit unsteady Kutta condition with a shedding panel of fixed
  geometry, which keeps the influence matrix constant in time;
\item a desingularized kernel applied to each vortex \emph{line} of the wake
  lattice, so that lines shared by two rings cancel exactly, and applied also
  to the body rings when the wake nodes are convected;
\item a bordered LU factorization of the singular Neumann problem of a closed
  body, which reproduces the minimum-norm (pseudo-inverse) solution at the
  cost of an LU factorization, extended to several bodies;
\item interior velocity conditions for ring-shaped bodies (box wings), which
  fix a nearly singular mode of their Neumann problem;
\item the detection of the separation lines from the grid connectivity,
  covering folded trailing edges (wings with tips, flat tip caps) and closed
  trailing edges (box and ring wings);
\item far-field loads from the circulation (Trefftz-plane induced drag,
  Kutta--Joukowski lift and propeller loads), which avoid the first-order
  error of the pressure of this discretization;
\item the unsteady Bernoulli equation written for the potential of the
  doublet method, including that of the discrete interior flow, and for
  rigidly moving bodies;
\item the image method for symmetric configurations, bodies without wake
  and a treatment of the collision of a wake with a body;
\item a strip-wise integral boundary layer coupled through the transpiration
  velocity, for the displacement effect and the profile drag.
\end{itemize}
Some of these choices were made after the analysis of an earlier
implementation of the method, whose lift did not converge under grid
refinement; the causes and the evidence are summarized in
\ref{app:earlier}.

\section{Mathematical model}
\label{sec:model}

\subsection{Potential flow and boundary conditions}

In a frame fixed to the body (standard aerodynamic body axes: $x$
downstream, $y$ spanwise, $z$ up), the incompressible irrotational flow
outside the body and its wake is $\V=\V_\infty+\nabla\phi$, with
$\nabla^2\phi=0$. The angle of attack $\alpha$ and the sideslip $\beta$ of the
body, rotated rigidly, define the free stream,
$\V_\infty=|\V_\infty|(\cos\alpha\cos\beta,\sin\beta,\sin\alpha\cos\beta)$;
local incidences of the parts of the body are part of its geometry.
On the body surface $S_B$ the flow tangency condition, modified by the
displacement of the boundary layer, reads
\begin{equation}
\V\cdot\n = v_n \quad \text{on } S_B,
\label{eq:bc}
\end{equation}
where $\n$ is the outward normal and $v_n$ the transpiration velocity
(Section~\ref{sec:bl}); $v_n=0$ in inviscid computations. The perturbation
vanishes at infinity. Vorticity is confined to the body surface and to the
wake sheets $S_W$ shed from the separation lines; the wake is force free and
moves with the local velocity, and the circulation of every wake element is
conserved (Kelvin, Helmholtz).

\subsection{Vortex-ring representation}

A doublet sheet of strength $\mu$ is equivalent to a vortex sheet of strength
$\bm\gamma=\n\times\nabla\mu$; a piecewise-constant doublet distribution on a
quadrilateral panel is therefore equivalent to a closed vortex ring of
circulation $\Gamma=\mu$ along the panel perimeter~\cite{katz2001}. The body and
the wake are both represented by vortex rings. On a closed surface, a uniform
$\Gamma$ induces no velocity: the interior Neumann problem determines $\Gamma$
up to a constant, and the discrete influence matrix is singular
(Section~\ref{sec:linsys}).

\subsection{Kutta condition}

At a separation line the flow must leave the surface smoothly. With vortex
rings, the discrete unsteady Kutta condition prescribes that the net vortex
on the separation edge vanishes~\cite{morino1974,katz2001}:
\begin{equation}
\Gamma_{w} = \Gamma_A - \Gamma_B ,
\label{eq:kutta}
\end{equation}
where $\Gamma_A$, $\Gamma_B$ are the rings on the two sides of the edge and
$\Gamma_w$ the first wake ring shed from it. The circulation shed at time $t$
is then convected with the wake.

\subsection{Loads}

The pressure coefficient follows from the Bernoulli equation. The steady
form $C_p=1-|\V|^2/|\V_\infty|^2$ is exact when the flow has become steady,
which is the case of the results of this paper unless stated otherwise;
optionally the unsteady term is included (Section~\ref{sec:rot}). The near-field force is the integral of the
pressure over the surface. For the induced drag the far-field (Trefftz-plane)
form is used~\cite{katz2001}:
\begin{equation}
D_i = -\frac{\rho}{2}\oint_{S_W\cap T} \Delta\phi\; \frac{\partial\phi}{\partial\nu}\,\mathrm d\ell ,
\label{eq:trefftz}
\end{equation}
where $T$ is a plane normal to $\V_\infty$ far downstream, $\Delta\phi=\Gamma$
the jump of potential across the wake trace and $\nu$ its normal.
Consistently with it, the far-field lift is the Kutta--Joukowski force of the
bound vortices of the shed rings on the separation lines,
$\mathbf F=\rho\sum_k\Gamma_k\,\V_\infty\times\Delta\mathbf l_k$; the pressure
lift converges to the same value (Section~\ref{sec:boxwing}).

\section{Numerical method}
\label{sec:method}

\subsection{Surface discretization}

The body is described by a structured grid of nodes $\x_{i,j}$,
$i=1,\dots,n_s+1$ (the line $i=n_s+1$ is the trailing edge),
$j=1,\dots,n_c+1$ (the line $j=n_c+1$ coincides with $j=1$). Each
quadrilateral panel carries a vortex ring along its perimeter. The
collocation point is the mean of the four vertices, the normal is the
normalized cross product of the diagonals and the area half its modulus.
Neighbouring panels, including those across the leading edge, the trailing
edge and the seam, are identified by matching the node coordinates; the
orientation of the normals is checked through the sign of the enclosed volume
$\frac13\sum_i(\x_i\cdot\n_i)A_i$.

\subsection{Biot--Savart kernel}

The velocity induced at $\x_P$ by a straight vortex segment from $\x_a$ to
$\x_b$ of unit circulation is~\cite{katz2001}
\begin{equation}
\q = \frac{1}{4\pi}\,
\frac{\mathbf r_1\times\mathbf r_2}{|\mathbf r_1\times\mathbf r_2|^2+\delta^2|\mathbf r_0|^2}\,
\mathbf r_0\cdot\left(\frac{\mathbf r_1}{|\mathbf r_1|}-\frac{\mathbf r_2}{|\mathbf r_2|}\right),
\label{eq:kernel}
\end{equation}
with $\mathbf r_0=\x_b-\x_a$, $\mathbf r_1=\x_P-\x_a$,
$\mathbf r_2=\x_P-\x_b$. For $\delta=0$ this is the exact kernel (set to zero
on the segment axis); for $\delta>0$ the velocity is regularized with a core
of radius $\delta$ and goes smoothly to zero on the axis, as in the
Rankine-type models used in free-vortex wake methods~\cite{vatistas1991,leishman2002}.
In the boundary condition the body rings use the exact kernel, which is
appropriate since the collocation points lie at a distance of order of the
panel size from the ring edges, and the exact normal influence is needed. The
wake uses $\delta=r_c|\V_\infty|\Delta t$, i.e.\ a core proportional to the
wake panel length ($r_c=0.5$). When the velocity of the wake nodes is
computed, instead, every line has the core $\delta$, the body rings included:
a wake node that comes close to a body would otherwise get a $1/r$ velocity.
This happens in particular with tip separation, where the tip sheets run
along the tip caps: with the exact body kernel their nodes reach velocities
of $36|\V_\infty|$ and the loads fluctuate irregularly in time
(Section~\ref{sec:mcalister}). The trailing-edge edges of the body rings and
the trailing-edge line of the wake, which coincide with opposite
circulations, keep the same kernel and still cancel.

\subsection{Consistency of the wake kernel}
\label{sec:consistency}

The wake is a structured lattice of rings; each interior vortex line is shared
by two rings with opposite orientation and carries the difference of their
circulations. If the two rings of a shared line are evaluated with different
kernels, the line does not cancel: it leaves a spurious vortex whose strength
is of the order of the circulation itself, not of its difference. This
happens naturally when the first wake row (which enters the influence matrix,
Section~\ref{sec:kutta}) is evaluated with the exact kernel and the following
rows (on the right-hand side) with the regularized one: the line between the
two rows, at a distance of order $|\V_\infty|\Delta t$ from the trailing edge,
then carries a spurious vortex of strength $\approx\Gamma_w$. In the present
method every line of the wake lattice has a single kernel, whatever ring it
belongs to: the trailing-edge line (which must cancel the trailing-edge
edges of the body rings exactly) uses $\delta=0$, all the other lines use
$\delta>0$. Section~\ref{sec:sens} shows that the inconsistent treatment
changes the steady lift by a factor up to 2.5 as the length of the first wake
row varies, and that the consistent one makes it independent of it.

The same property is exploited for efficiency: the wake and, when needed, the
body are evaluated as lists of vortex lines, each line once with its net
circulation, which halves the number of kernel evaluations with respect to the
sum over rings and gives the same result up to round-off.

\subsection{Wake and implicit unsteady Kutta condition}
\label{sec:kutta}

Each wake sheet is a lattice of nodes $\x^w_{s,r}$, $s=1,\dots,n_{sp}+1$
across the sheet and $r=1,\dots,n_w+1$ downstream, and rings $(k,r)$ between
node rows $r$ and $r+1$. Node row 1 is the separation line. Node row 2 is
placed at $\x^w_{s,1}+f\,\Delta t\,\V_\infty$ and is \emph{fixed}: the first
ring row (the \emph{Kutta rings}) has a constant geometry, and its
circulation is given by the Kutta condition~(\ref{eq:kutta}). Since the
Kutta rings are fixed, their influence can be included in the influence
matrix once and for all: the velocity induced by Kutta ring $k$ is added to
the column of panel $A_k$ and subtracted from the column of panel $B_k$. The
condition is then imposed implicitly at each time step. No velocity is ever
evaluated on the separation line itself, where it is singular: the first
wake row is placed along $\V_\infty$ (along the relative velocity for moving
bodies). An explicit condition, with the circulation of the preceding step,
and a first row placed with the velocity computed on the trailing edge are
causes of non-convergence (\ref{app:earlier}).

After the solution at time $t_n$ the velocity of the free nodes (rows
$2,\dots,n_w+1$) is computed from the body, the whole wake and $\V_\infty$,
and the wake is advanced with an explicit Euler step: node row $r$ moves to
row $r+1$ and ring row $r$ becomes row $r+1$, while rows 1 and 2 stay in
place and the new Kutta circulation is set by the next solution. Rows beyond
a prescribed number $n_{w,\max}$ may be dropped. The parameter $f$ ($=1$) sets
the length of the Kutta rings; the steady solution does not depend on it
(Section~\ref{sec:sens}). The time step may be tied to the grid,
$|\V_\infty|\Delta t=k\,h_{span}$, with $h_{span}$ the mean spanwise width of
the trailing-edge panels: for $k=1$ the wake panels are square, and the time
step is refined together with the body grid, as required for the convergence
of the tip sheets (Section~\ref{sec:sens}).

\subsection{Linear system}
\label{sec:linsys}

Imposing~(\ref{eq:bc}) at the $N$ collocation points gives
\begin{equation}
\mathsf A\,\bm\Gamma = \mathbf b,\qquad
A_{ij}=\n_i\cdot\q_j(\x_i),\qquad
b_i=-(\V_\infty+\V_w(\x_i))\cdot\n_i+v_{n,i},
\end{equation}
where $\q_j$ includes the Kutta rings attached to panel $j$ and $\V_w$ is the
velocity of the free wake. For a closed body $\mathsf A\mathbf 1=\mathbf 0$ and
the matrix has rank $N-1$. Rather than computing the pseudo-inverse by a
singular value decomposition, which is expensive and parallelizes poorly,
the system is bordered with the gauge condition $\sum_j\Gamma_j=0$:
\begin{equation}
\mathsf M\begin{pmatrix}\bm\Gamma\\ \lambda\end{pmatrix}=
\begin{pmatrix}\mathsf A & s\mathbf 1\\ s\mathbf 1^T & 0\end{pmatrix}
\begin{pmatrix}\bm\Gamma\\ \lambda\end{pmatrix}=
\begin{pmatrix}\hat{\mathbf b}\\ 0\end{pmatrix},
\label{eq:bordered}
\end{equation}
where $s$ is the mean diagonal of $\mathsf A$. $\mathsf M$ is regular and is
factorized once (LAPACK \texttt{dgetrf}~\cite{lapack1999}). Because of the
discretization, $\mathbf b$ has a small component along the left null vector
$\mathbf w$ of $\mathsf A$ ($\mathsf A^T\mathbf w=\mathbf 0$). The vector
$\mathbf w$ is obtained from the transposed system
$\mathsf M^T(\mathbf w,\mu)^T=(\mathbf 0,1)^T$, which gives
$\mathsf A^T\mathbf w=-\mu s\mathbf 1$ and, since $\mathbf 1$ is not in the
range of $\mathsf A^T$ (orthogonal to the null space $\mathbf 1$ of
$\mathsf A$) unless $\mu=0$, $\mathsf A^T\mathbf w=\mathbf 0$. The
right-hand side is then projected on the range of $\mathsf A$,
$\hat{\mathbf b}=\mathbf b-(\mathbf w\cdot\mathbf b/\mathbf w\cdot\mathbf w)\,\mathbf w$,
so that $\lambda=0$, $\mathsf A\bm\Gamma=\hat{\mathbf b}$ and
$\mathbf 1^T\bm\Gamma=0$: $\bm\Gamma$ is exactly the minimum-norm solution
$\mathsf A^+\mathbf b$. The relative size $|\mathbf w\cdot\mathbf b|/|\mathbf b|$
is a measure of the discretization error of the closed surface (about
$5\times10^{-3}$ in the cases of this paper). For an open surface the plain
LU factorization of $\mathsf A$ is used.

\subsection{Several bodies and ring-shaped bodies}
\label{sec:multi}

With several bodies the unknowns of all of them form one system, so that
every body and every wake acts on all the others. A uniform circulation on a
closed body induces no velocity anywhere (every edge is shared by two of its
rings): $\mathsf A$ has one null vector $\mathbf 1_c$ per closed body $c$, the
system~(\ref{eq:bordered}) is bordered with one gauge condition
$\mathbf 1_c^T\bm\Gamma=0$ per closed body, and the left null vectors,
obtained from the transposed system with the right-hand sides
$(\mathbf 0,\mathbf e_c)$ and orthonormalized, are all removed from
$\mathbf b$.

On a ring-shaped body (closed body with a closed trailing edge, e.g.\ a box
wing) the Neumann problem has a further, nearly singular direction: a
circulation level that differs between the parts of the ring and changes
along its thin parts (the links of a box wing). It is equivalent to vortex
rings wound around the link, like a solenoid: it drives a strong flow inside
the link but changes the normal velocity outside only slightly, and the
smaller the cross-section of the link, the less. Small discretization errors
can then excite it, and produce a spurious spanwise flow along the links,
with the sectional circulation almost unaffected but $C_p$ of $-10$ to $-250$
on the whole link; with a fixed panel number it appeared in about one third
of random box wings with short-chord links. The exact solution, however, has
zero total velocity inside the body: the interior flow is harmonic, with zero
normal velocity on the wall and a single-valued potential. For every
ring-shaped body the condition
\begin{equation}
\V_\infty+\sum_j\Gamma_j\,\q_j(\x_k)+\V_w(\x_k)=\mathbf 0
\end{equation}
is therefore imposed at one interior point $\x_k$ per panel column (the
mid-thickness point of the thickest station), three equations per point, with
the kernels of the boundary condition. The stacked system (Neumann rows,
gauge rows, interior rows with the weight $w=0.01$) is solved in the
least-squares sense by a QR factorization computed once. The small weight
leaves the boundary condition practically exact and only fixes the nearly
singular direction: with $w=1$ the least-squares compromise spoiled the
trailing-edge panels, with $w\le0.003$ the mode was no longer controlled.
Section~\ref{sec:boxwing} shows the effect.

\subsection{Symmetry and non-lifting bodies}
\label{sec:sym}

A configuration symmetric about the $x$--$z$ plane (with $\beta=0$) is
computed on its half $y\ge0$ by the image method: every vortex segment has
its mirror image, carrying the mirrored vorticity (a pseudovector: the
mirrored ring is traversed with the opposite circulation). The velocity that
the image of a segment induces at $\x$ is the mirror of the velocity that the
segment induces at the mirror point,
$\q_{img}(\x)=\mathsf M\,\q(\mathsf M\x)$ with $\mathsf M=\mathrm{diag}(1,-1,1)$;
applied in the segment kernel, this includes the images in every
Biot--Savart evaluation (matrix, right-hand side, wake, interior points). A
body cut by the plane is closed with its image: its edges on $y=0$ are not
free edges, and its uniform circulation, with that of the image, still
induces no velocity, so it keeps its gauge condition; in the surface
vorticity these edges face the mirror panel, which has the same
circulation. The end of a trailing-edge sheet on the plane is not a tip, and
the Trefftz plane includes the image trailing vortices. The unknowns are
halved, the cost of the matrix is halved and that of the factorization
divided by eight.

A body sheds a wake only where the connectivity finds a trailing edge; the
wake of any body can also be switched off, which makes it non-lifting (a
fuselage or a nacelle between lifting wings). Bodies are not intersected: a
wing ends at the side of a fuselage, with a small gap.

\subsection{Rotating bodies and unsteady pressure}
\label{sec:rot}

Any body can be given a rigid rotation $\bm\Omega=\omega\,\mathbf e$ about an axis
through $\x_0$ (propeller or rotor blades, one body each). The computation is
made in the frame of the free stream: at every step the nodes are rotated,
the boundary condition becomes $(\V-\V_b)\cdot\n=0$ with
$\V_b=\bm\Omega\times(\x-\x_0)$, the Kutta rings are attached to the moved
separation lines and shed along $\V_\infty-\V_b$, and the free wake is
convected with the absolute velocity. When all the bodies rotate together
and $\V_\infty$ is parallel to the axis, the configuration is rigid in its own
frame: the influence matrix is built and factorized once and the velocity
matrices are rotated; otherwise (relative motion between bodies, oblique
flow, or a rotating body with its mirror image) it is rebuilt at every step.

On the surface the outer potential is $\Phi=\mu+\phi_i$, where $\mu$ is the
jump of the potential (the ring circulation) and $\phi_i$ the potential of the
interior flow on the wall. Following a body point, with $p=p_\infty$ far
away, the Bernoulli equation gives
\begin{equation}
C_p=\frac{|\V_\infty|^2+|\V_b|^2-|\V-\V_b|^2-2\,\mathrm d\Phi/\mathrm dt}{V_{ref}^2},
\qquad \frac{\mathrm d\Phi}{\mathrm dt}=\frac{\mathrm d\mu}{\mathrm dt}+\frac{\mathrm d\phi_i}{\mathrm dt},
\label{eq:cpunst}
\end{equation}
up to a uniform term, which gives no force on a closed body. For the exact
solution the interior flow satisfies the Neumann data $\V_b\cdot\n$ and its
potential is fixed in the frame of a rigidly moving body, so that
$\mathrm d\phi_i/\mathrm dt$ is uniform. The discrete Neumann solution, however, has
an interior flow of order $h$ (Section~\ref{sec:boxwing}), which changes in
time with the flow around the body: on a fuselage in the growing flow field
of a wing, $\mathrm d\mu/\mathrm dt$ alone produced a spurious pressure gradient
along the body ($C_p$ from $-2$ at the nose to $+1$ at the tail). The interior
velocity on the wall is known exactly in terms of the solution,
$\V-\bm\gamma\times\n$, and its potential is single-valued (the jump across
the wake is in $\mu$); $\phi_i$ is therefore reconstructed on every body by
least squares over the edges between panels,
$\phi_{i,q}-\phi_{i,p}=\tfrac12(\V_{t,p}+\V_{t,q})\cdot(\x_q-\x_p)$, by a few
conjugate-gradient iterations per step from the previous solution. Both
derivatives are backward differences following the panel. Without motion
Eq.~(\ref{eq:cpunst}) is the classical unsteady correction
$-2\,\partial\Phi/\partial t$; when the flow is steady in the frame of the body it
reduces to $(|\V_\infty-\V_b|^2-|\V-\V_b|^2)/V_{ref}^2$ only if $\V_\infty\cdot\V_b$
is zero or uniform (axial flow, translation).

Because of the first-order error of the pressure on the chordwise panel size
(Section~\ref{sec:boxwing}), the blade loads are also computed from the
circulation, as in lifting-line theory: each Kutta ring $k$ carries
$\mathbf F_k=\rho\,\Gamma_k\,\mathbf W_k\times\Delta\mathbf l_k$ on its bound vortex, the
separation line, with $\mathbf W_k$ the relative velocity plus the velocity
induced by all the wakes and all the other bodies, without the bound
vorticity of the blade itself. The trailing legs start at the trailing edge,
where their induced velocity is half the far-wake value, as at the lifting
line.

\subsection{Wake--body collision}
\label{sec:coll}

A free wake convected with an explicit scheme can cross a body, for
example a propeller slipstream reaching a wing: the continuous flow would
deflect it, but a node can jump through the surface in one step. When this
matters, every free node inside the bounding box of a body is compared, after
the convection, with the nearest panel of that body: if its projection falls
on the panel and it lies on the inner side, or closer than a clearance of a
quarter of the panel size, it is moved along the normal to that clearance.
The Kutta rings, attached to the separation lines, are excluded, and the
projection test leaves the nodes just behind a trailing edge untouched. The
search costs the number of nodes in the boxes times the panels of the body
and is switched on only when needed.

\subsection{Surface velocity}

The total velocity at the collocation point $i$, on the outer side of the
vortex sheet, is
\begin{equation}
\V_i=\V_\infty+\sum_j\Gamma_j\,\q_j(\x_i)+\V_w(\x_i)+\tfrac12\,\bm\gamma_i\times\n_i ,
\end{equation}
where the sum (the principal value of the body and Kutta rings) is evaluated
with three precomputed velocity influence matrices, and $\bm\gamma_i$ is the
local surface vorticity. The vortex segment on edge $k$ of panel $i$ carries
$\Gamma_i-\Gamma_{nb(k)}$ and is spread over the two panels:
\begin{equation}
\bm\gamma_i=\sum_{k=1}^4\frac{\Gamma_i-\Gamma_{nb(k)}}{A_i+A_{nb(k)}}\,\mathbf e_k ,
\end{equation}
with $\mathbf e_k$ the edge vector; this is a central difference of $\Gamma$
along the surface. On a separation edge the net circulation vanishes by the
Kutta condition, and on a free edge there is no neighbour: that edge is
skipped and the opposite one counts twice (a one-sided difference).

\subsection{Separation lines}

The separation lines are derived from the connectivity of the panels
$i=n_s$ across their downstream edge. If the partners lie on the line
$i=1$, the section is wrapped around the trailing edge (closed trailing edge,
box or ring wing): the trailing-edge sheet covers the whole line and has no
free edges, and there are no tips. If the partners are other trailing-edge
panels, the trailing-edge line folds back on itself at two places, where two
consecutive panels are partners, possibly separated by a run of zero-length
edges (a flat tip cap): the sheet is shed between the folds, and the fold
edges are the tips, from which two additional sheets are shed when tip
separation is requested. Trailing-edge edges of zero length (e.g.\ at the
pole of a body of revolution) shed rings of zero circulation.

\subsection{Integral boundary layer}
\label{sec:bl}

The body is divided into chordwise strips (for a folded trailing edge a panel
column joined through the leading edge to its partner column; for a closed
trailing edge each column). On each strip the edge velocity $U_e$ is the
computed surface speed, and the boundary layer is integrated from the
stagnation point (minimum $U_e$) to the two trailing edges. Laminar flow is
computed with Thwaites' method~\cite{thwaites1949},
\begin{equation}
\theta^2=\frac{0.45\,\nu}{U_e^6}\int_0^sU_e^5\,\mathrm ds,\qquad
\lambda=\frac{\theta^2}{\nu}\frac{\mathrm dU_e}{\mathrm ds},\qquad
c_f=\frac{2\nu\,\ell(\lambda)}{U_e\theta},
\end{equation}
with the correlations of Cebeci and Bradshaw~\cite{cebeci1977}:
$\ell=0.22+1.57\lambda-1.8\lambda^2$, $H=2.61-3.75\lambda+5.24\lambda^2$
for $\lambda\ge0$, and $\ell=0.22+1.402\lambda+0.018\lambda/(\lambda+0.107)$,
$H=2.088+0.0731/(\lambda+0.14)$ for $\lambda<0$. Transition occurs when
Michel's criterion~\cite{michel1951}
$Re_\theta\ge1.174\,(1+22400/Re_x)\,Re_x^{0.46}$ is met or when the laminar
layer separates ($\lambda\le-0.09$). The turbulent layer is computed with
Head's entrainment method~\cite{head1958},
\begin{equation}
\frac{\mathrm d\theta}{\mathrm ds}=\frac{c_f}{2}-(H+2)\frac{\theta}{U_e}\frac{\mathrm dU_e}{\mathrm ds},\qquad
\frac{\mathrm d(U_e\theta H_1)}{\mathrm ds}=0.0306\,U_e\,(H_1-3)^{-0.6169},
\end{equation}
with $H_1=3.3+0.8234(H-1.1)^{-1.287}$ for $H\le1.6$,
$H_1=3.3+1.5501(H-0.6778)^{-3.064}$ for $H>1.6$, the Ludwieg--Tillmann skin
friction $c_f=0.246\cdot10^{-0.678H}Re_\theta^{-0.268}$~\cite{ludwieg1949},
initial shape factor $H=1.4$, and separation when $H>2.4$. The integration
of a side also stops when the edge velocity falls below $0.02|\V_\infty|$
(stagnant flow) or $\theta$ exceeds 5\% of the strip length; in all cases the
last valid values are frozen downstream. The profile drag of each strip is
given at the trailing edge by the Squire--Young formula~\cite{squire1938},
per unit span and per side, $D'/q=2\theta\,(U_e/|\V_\infty|)^{(H+5)/2}$,
with $H$ limited to 2.5.

The displacement effect enters the boundary condition~(\ref{eq:bc}) through
the transpiration velocity $v_n=\mathrm d(U_e\delta^*)/\mathrm ds$~\cite{lighthill1958},
which is under-relaxed (factor 0.5) and applied at the next time step: the
viscous--inviscid coupling converges together with the time marching towards
the steady state. The friction force is the integral of
$c_f(U_e/|\V_\infty|)^2$ along the local velocity direction. The boundary
layer is quasi-steady and two-dimensional along each strip: cross-flow,
wake displacement and stall are not modelled.

\subsection{Induced drag in the Trefftz plane}

Equation~(\ref{eq:trefftz}) is discretized with the trailing vortices that
leave the separation lines along $\V_\infty$: at node $s$ of the separation
line the vortex has circulation $\Gamma_{s-1,1}-\Gamma_{s,1}$ (circulations
of the Kutta rings), and its trace in the Trefftz plane is the projection of
the node. With $\mathbf u_k$ the induced velocity (two-dimensional point
vortices) at the midpoint of the trace element $k$, of length $\ell_k$ and
normal $\bm\nu_k$,
$D_i=-\frac{\rho}{2}\sum_k\Gamma_{k,1}(\mathbf u_k\cdot\bm\nu_k)\ell_k$. This is
reliable when the wake is shed from the trailing edge; the tip sheets leave
a line almost parallel to $\V_\infty$, whose trace is degenerate, and for
them the formula is only approximate.

\subsection{Implementation}

BBun is written in Fortran~90. The influence
matrices are built in parallel with OpenMP~\cite{openmp2018}, and all the
Biot--Savart evaluations of the time loop (right-hand side, wake velocity)
and the boundary layer are parallel loops. The LU factorization and the
matrix--vector products use OpenBLAS~\cite{openblas2012}. The results are
written as legacy VTK files~\cite{vtk} in double precision; the body and the
wake are separate files, and any output step is a complete restart point.
Several bodies are given as several structured grids in one input file: every
body has its own connectivity, separation lines, wake sheets and boundary-layer
strips, while the unknowns of all the bodies form one linear system, so that
all the bodies and all the wakes interact; every closed body adds its own
gauge condition to the bordered system.

\section{Scope and limitations}
\label{sec:scope}

The model is incompressible potential flow with thin free vortex sheets and a
quasi-steady, strip-wise boundary layer. It is meant for attached flow at
moderate incidence, for the comparison of configurations and for the
interaction of lifting bodies, wakes and propellers, not for the prediction
of separation or stall. Table~\ref{tab:limits} lists the assumptions and the
known errors of the discretization, with the section where each is
quantified; Section~\ref{sec:concl} gives the resulting recommendations for
use.

\begin{table}[t]
\centering
\caption{Assumptions and known limitations of the model and of its
discretization.}
\label{tab:limits}
\small
\begin{tabular}{@{}p{3.4cm}p{8.6cm}p{2.2cm}@{}}
\toprule
item & limitation & see\\
\midrule
flow model & incompressible, inviscid outside the boundary layer; no
compressibility correction (the tip Mach number of propellers is not
accounted for) & \\
surface discretization & constant-strength vortex rings with the Neumann
condition: first-order error of the surface velocity in the chordwise
panel size, a nearly uniform speed deficit with an $O(h)$ interior flow;
pressure lift about $-130/n$\,\% with $n$ panels per side, while the
circulation converges fast & Sec.~\ref{sec:accuracy}\\
thin sections & a symmetric suction peak at the trailing edge
($C_p\approx-1$ over the last 3\% of the chord for $t/c=0.06$) & Sec.~\ref{sec:accuracy}\\
near-field drag & the drag of the pressure integration is 0.005--0.01 too
high; the drag is taken from the Trefftz plane and the boundary layer &
Sec.~\ref{sec:mcalister}\\
pressure & steady Bernoulli equation by default; the unsteady term is an
option & Sec.~\ref{sec:rot}\\
wake & thin sheets of vortex rings with a core, explicit Euler convection,
no viscous diffusion; tip separation only approximate for the induced drag &
Sec.~\ref{sec:sens}\\
wake--body interaction & no deformation of a wake around a body beyond the
push-out of the collision treatment (optional) & Sec.~\ref{sec:coll}\\
geometry & structured grids, one per body; bodies are not intersected (a
wing ends at the side of a fuselage, with a small gap and a root vortex) &
Sec.~\ref{sec:sym}\\
boundary layer & quasi-steady, two-dimensional along chordwise strips: no
cross-flow, no rotational effects on blades, no wake displacement, no stall
& Sec.~\ref{sec:bl}\\
rotating bodies & rigid rotation about a fixed axis; loads from the
circulation are quasi-steady; no induced drag for helical wakes; hover not
validated & Sec.~\ref{sec:rot}\\
validation data & wind-tunnel data without wall corrections; no
measurements for the unsteady and the propeller cases (theory only) &
Sec.~\ref{sec:results}\\
\bottomrule
\end{tabular}
\end{table}

\section{Verification and validation}
\label{sec:results}

The results are presented in full, since they define the range of validity
of the code for its applications. They comprise the verification of the
discretization (a sphere, the sensitivity to the parameters and the grid
convergence, the two-dimensional limit, independent vortex-lattice
computations of single and interacting wings, the symmetry and the motion of
bodies), the validation against experiments (the NACA~0015 wing of McAlister
and Takahashi, the NACA~0012 profile drag, the biplane of Knight and Noyes),
the comparison with classical theory (Prandtl's best wing system, Wagner's
function, blade-element momentum theory), the accuracy of the pressure loads,
an application to a complete aircraft and the computational performance.
All the computations use closed trailing edges, the parameters $f=1$ and
$r_c=0.5$, and are marched in time from an impulsive start until the loads
are steady unless stated otherwise. The coefficients are referred to the
planform area unless stated otherwise.

\subsection{Sphere}

\begin{figure}[t]
\centering
\includegraphics[width=0.55\textwidth]{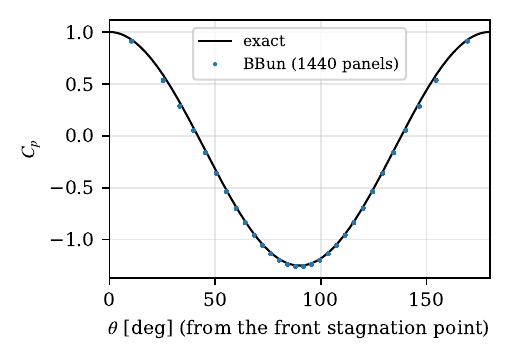}
\caption{Pressure coefficient on a sphere, 1440 panels.}
\label{fig:sphere}
\end{figure}

The potential flow around a sphere, $C_p=1-\tfrac94\sin^2\theta$, verifies
the body discretization, the treatment of the null space and the surface
velocity (no wake is shed: the trailing-edge edges at the pole have zero
length). Figure~\ref{fig:sphere} shows the solution with 1440 panels;
Table~\ref{tab:sphere} the error for three grids. The RMS error decreases
with an order of about 0.7--0.8 with respect to the panel size; the maximum error
is located at the poles, where the panels degenerate into triangles.

\begin{table}[t]
\centering
\caption{Error of $C_p$ on the sphere.}
\label{tab:sphere}
\begin{tabular}{rcc}
\toprule
panels & max $|C_p-C_{p,exact}|$ & RMS\\
\midrule
360 & 0.072 & 0.033\\
1440 & 0.051 & 0.021\\
5760 & 0.032 & 0.012\\
\bottomrule
\end{tabular}
\end{table}

\subsection{Parameter sensitivity and grid convergence}
\label{sec:sens}

The effect of the wake kernel consistency (Section~\ref{sec:consistency}) is
shown in Table~\ref{tab:sens}, for a rectangular wing ($t/c=0.12$ at the root,
thickness vanishing at the tips, span 5 chords) at $\alpha=26.6^\circ$
without tip separation. With a frozen wake (convected with $\V_\infty$) the
steady lift must not depend on the length $f|\V_\infty|\Delta t$ of the Kutta
rings, since all the wake rings then have the same circulation and only the
splitting of the sheet changes. With the inconsistent kernel the lift varies
by a factor 2.5 for $f$ between 1 and 0.25; with the consistent one it varies
by 0.2\%. With the free wake the residual dependence on $f$ and $r_c$ is
1.6\% and 1.3\% during the transient ($t=3$ chords).

\begin{table}[t]
\centering
\caption{Normal force $F_z/(qS_{wet})$ of a wing at $\alpha=26.6^\circ$
(1152 panels) for different lengths of the Kutta rings ($f$) and core radii
($r_c$), in units of $|\V_\infty|\Delta t$.}
\label{tab:sens}
\begin{tabular}{lccc}
\toprule
case & $f=1$ & $f=0.5$ & $f=0.25$\\
\midrule
frozen wake, inconsistent kernel, $t=15$ & 0.749 & 0.526 & 0.300\\
frozen wake, consistent kernel, $t=6$ & 0.865 & 0.864 & 0.863\\
free wake, consistent kernel, $t=3$ & 0.819 & 0.810 & 0.806\\
\midrule
 & $r_c=0.25$ & $r_c=0.5$ & $r_c=1$\\
\midrule
free wake, $f=1$, $t=3$ & 0.816 & 0.819 & 0.826\\
\bottomrule
\end{tabular}
\end{table}

\begin{figure}[t]
\centering
\includegraphics[width=\textwidth]{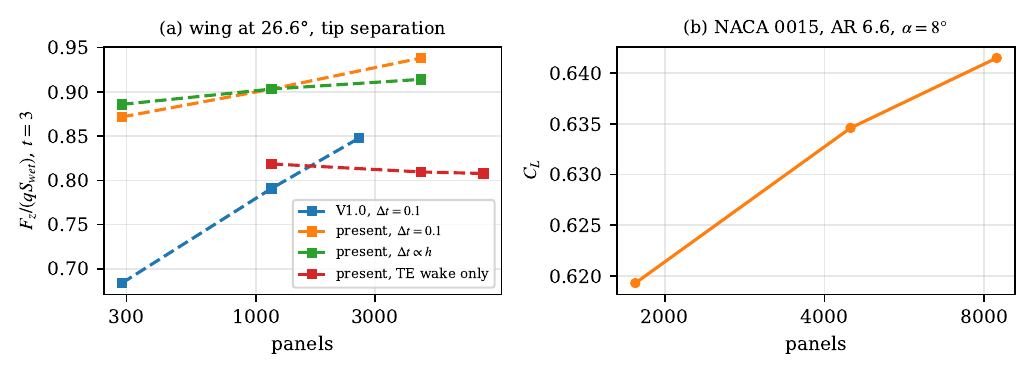}
\caption{Grid convergence. (a) Wing at $\alpha=26.6^\circ$ with tip
separation: earlier implementation (\ref{app:earlier}) and present
method with fixed and refined time step, and with the trailing-edge wake only. (b) NACA~0015 wing of
aspect ratio 6.6 at $\alpha=8^\circ$ (inviscid).}
\label{fig:conv}
\end{figure}

Figure~\ref{fig:conv}(a) compares the grid refinement of the earlier
implementation (\ref{app:earlier}) and of the present method on
the same wing with tip separation. The earlier implementation does not
converge: the force grows by 16\% and 7\% at each refinement. The present method with only the trailing-edge wake converges (changes of
$-1.1\%$ and $-0.2\%$). With tip separation and a fixed time step the result
still drifts by about 4\% per refinement: the tip sheet is resolved only at
intervals $|\V_\infty|\Delta t$ in the downstream direction, independent of
the body grid. Refining the time step together with the grid
($\Delta t\propto h$) restores the convergence ($+2.0\%$, $+1.2\%$).
Figure~\ref{fig:conv}(b) shows the convergence of the NACA~0015 wing of
Section~\ref{sec:mcalister}.

\subsection{Two-dimensional limit and three-dimensional lift slope}

The body discretization and the Kutta condition are verified in the
two-dimensional limit. An independent two-dimensional panel method (linear
vortex panels~\cite{katz2001}) gives for the NACA~0015 section with closed
trailing edge the inviscid lift slope $a_0=7.05$\,rad$^{-1}$, converged with
the number of panels and in agreement with $2\pi(1+0.77\,t/c)$. For a wing of
aspect ratio 40 at $\alpha=4^\circ$, BBun gives at mid span
$c_l=0.456$ at $t=30$ chords; lifting-line theory with $a_0=7.05$ gives
0.466 for the steady state and about 0.457 when the residual transient at
$t=30$ is accounted for. The agreement is within 1\%.

In three dimensions, the lift slope of the rectangular wing of aspect ratio
6.576 with square tips was compared with an independent planar
vortex-lattice method for the flat plate, which converges to
$C_{L\alpha}=4.36$\,rad$^{-1}$. BBun gives 4.36\,rad$^{-1}$ for a thin
NACA~0006 wing and 4.60\,rad$^{-1}$ for the NACA~0015 wing (the thickness
effect, $+5.5\%$), at $t=30$ chords. These are pressure lifts with 16
panels on each side of the section, about 4\% below the far-field lift, to
which they converge at first order in the chordwise panel size
(Section~\ref{sec:boxwing}).

\subsection{NACA 0015 wing: McAlister and Takahashi}
\label{sec:mcalister}

McAlister and Takahashi~\cite{mcalister1991} measured the surface pressure on
a rectangular NACA~0015 semispan wing with a square tip, mounted on an image
plane (full-wing aspect ratio 6.6), at chord Reynolds numbers of
$1$--$3\times10^6$; integrated loads are available for $2^\circ\le\alpha\le14^\circ$,
and the section loads along the span at 12 stations. The drag is obtained
from the pressures and therefore contains the induced and the viscous form
drag, not the friction. The data used here were digitized from the figures
of the report (marker centroids at 300\,dpi; uncertainty about $\pm0.02$ in
$C_L$ and $\pm0.002$ in $C_D$), for $Re=1.5$--$2.5\times10^6$. They are not
corrected for the interference of the wind-tunnel walls: the report states
it, and numerical studies of this case include the tunnel walls in the
computational domain~\cite{nalu}.

\begin{figure}[t]
\centering
\includegraphics[width=0.85\textwidth]{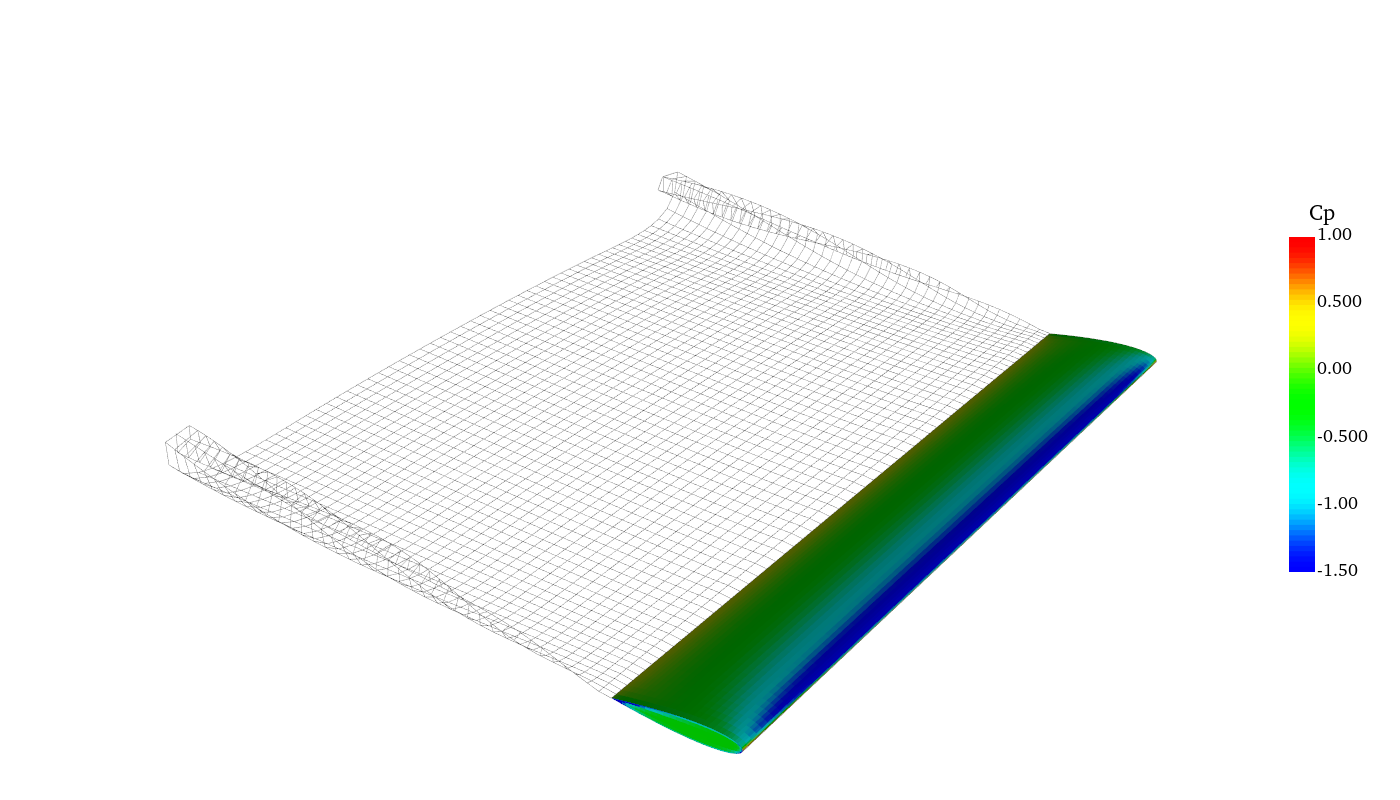}
\caption{NACA~0015 wing, $\alpha=8^\circ$, $Re=1.5\times10^6$: pressure
coefficient on the body and first four chords of the free wake.}
\label{fig:wing}
\end{figure}

The wing (chord 1, span 6.576, square tips closed by flat caps) is discretized
with 32 panels on each side of the section (cosine spacing), 32 panels on
each half span (clustered at the tips) and 6 panels across each cap, i.e.\
4480 panels; $\Delta t=0.1$, 150 steps (15 chords), whole wake retained. The
viscous computations use $Re=1.5\times10^6$ with free transition.
Figure~\ref{fig:wing} shows the pressure on the wing and the wake at
$\alpha=8^\circ$.

\begin{figure}[t]
\centering
\includegraphics[width=\textwidth]{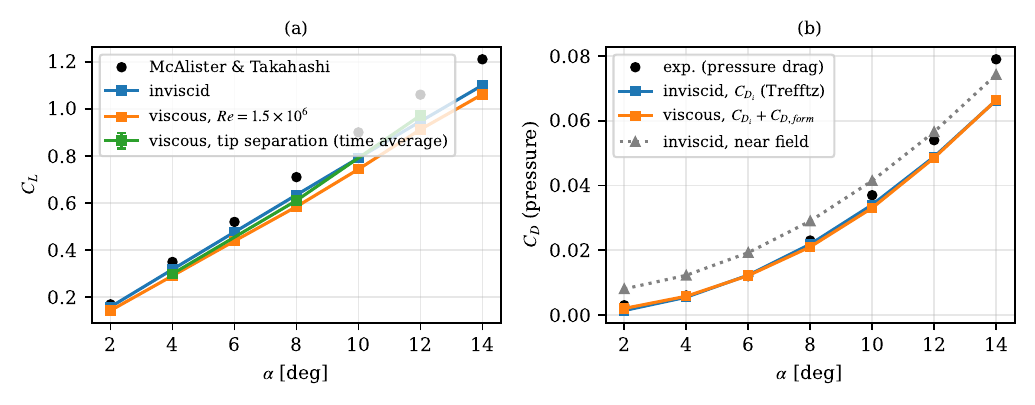}
\caption{NACA~0015 wing (AR 6.6, square tip). (a) Lift coefficient.
(b) Pressure drag: experiment, Trefftz-plane induced drag (inviscid),
induced plus form drag (viscous), and near-field drag from the surface
pressure (inviscid).}
\label{fig:mca}
\end{figure}

\begin{figure}[t]
\centering
\includegraphics[width=\textwidth]{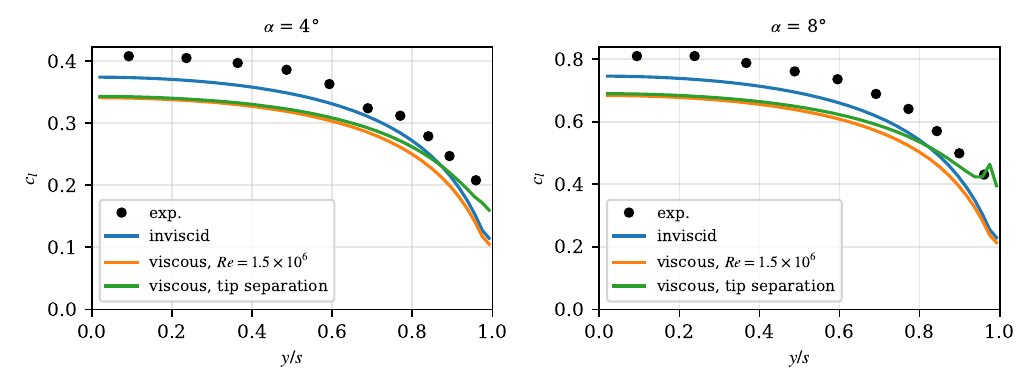}
\caption{Section lift along the span of the NACA~0015 wing, $\alpha=4^\circ$
and $8^\circ$; $s$ is the semispan.}
\label{fig:span}
\end{figure}

\paragraph{Lift}
Figure~\ref{fig:mca}(a) and Table~\ref{tab:mca} show the lift. The inviscid
lift is 6--12\% below the experiment, the viscous one 12--18\% below, the
boundary-layer displacement reducing the lift by 8--10\% up to $8^\circ$
and by 4--6\% above, where the separated rear strips are frozen. The
measured slope,
5.16\,rad$^{-1}$, is itself about 12\% above the potential-flow slope of the
planform given by the independent vortex-lattice method and by BBun
(4.36--4.60\,rad$^{-1}$). An upwash induced by closed tunnel walls,
$\Delta\alpha=\delta(S/C)C_L$~\cite{barlow1999}, with model-to-tunnel area
ratio $S/C\approx0.14$ and $\delta\approx0.1$--$0.2$, increases the measured
slope by the factor $1/(1-\delta(S/C)C_{L\alpha})\approx1.07$--$1.15$, which
is consistent with the difference. The comparison of the lift with this
dataset is therefore limited to about 10\%. The shape of the spanwise
loading (Fig.~\ref{fig:span}) is reproduced; with tip separation the lift
near the tip increases, as in the experiment, because of the tip vortex that
forms on the square edge; the same effect raises the integrated lift of the
tip-separation series in Fig.~\ref{fig:mca}(a) (by 3--7\% with respect to
the trailing-edge wake alone). The tip-sheet flow is regular: the loads
fluctuate by $\pm0.15\%$ at $8^\circ$ and $\pm0.25\%$ at $12^\circ$ (standard
deviation over the last five chords), and two computations that differ only
by round-off (the same grid with the coordinates perturbed by a relative
$10^{-12}$) coincide to $10^{-6}$ at $8^\circ$ and to $10^{-3}$ at $12^\circ$
over the whole run. With the exact kernel for the body rings at the wake
nodes, the tip-sheet nodes passing close to the tip caps got
velocities up to $36|\V_\infty|$, and the same comparison diverged after
20--28 steps, with fluctuations of $\pm1\%$ at $8^\circ$ and $\pm3$--$6\%$ at
$12^\circ$ and a mean lift 2\% ($8^\circ$) to 8\% ($12^\circ$) higher. The
values of this series are time averages over the last five chords.

\begin{table}[t]
\centering
\caption{NACA~0015 wing (AR 6.6): lift and pressure drag.}
\label{tab:mca}
\begin{tabular}{r ccc ccc}
\toprule
& \multicolumn{3}{c}{$C_L$} & \multicolumn{3}{c}{$C_D$ (pressure)}\\
\cmidrule(lr){2-4}\cmidrule(lr){5-7}
$\alpha$ & inviscid & viscous & exp. & $C_{D_i}$, inviscid & $C_{D_i}+C_{D,form}$ & exp.\\
\midrule
 2 & 0.159 & 0.144 & 0.17 & 0.0014 & 0.0020 & 0.003\\
 4 & 0.318 & 0.291 & 0.35 & 0.0055 & 0.0058 & 0.006\\
 6 & 0.477 & 0.439 & 0.52 & 0.0123 & 0.0122 & 0.012\\
 8 & 0.635 & 0.585 & 0.71 & 0.0218 & 0.0210 & 0.023\\
10 & 0.791 & 0.743 & 0.90 & 0.0341 & 0.0331 & 0.037\\
12 & 0.946 & 0.912 & 1.06 & 0.0489 & 0.0486 & 0.054\\
14 & 1.100 & 1.061 & 1.21 & 0.0663 & 0.0664 & 0.079\\
\bottomrule
\end{tabular}
\end{table}

\paragraph{Drag}
The induced drag computed in the Trefftz plane, and the induced plus form
drag of the viscous computations ($C_{D_i}+C_{D,prof}-C_{D,fric}$), agree with
the measured pressure drag within about 10\% up to $12^\circ$
(Fig.~\ref{fig:mca}(b), Table~\ref{tab:mca}). The span efficiency
$e=L^2/(\pi q b^2 D_i)$ computed with the far-field (Kutta--Joukowski) lift
of the shed circulation is 0.99 at $8^\circ$, against 0.98 from the
vortex-lattice method; with the pressure lift it would be 0.89, since on this
grid the pressure lift is 5\% below the far-field lift
(Section~\ref{sec:boxwing}). At
$14^\circ$ the experiment shows the onset of the drag growth associated with
separation, which the model does not describe. The near-field drag from the
integration of the surface pressure is 0.005--0.009 higher than the
Trefftz-plane value: the pressure integration is not accurate enough for
drag, and the far-field value should be used. With tip separation the
Trefftz-plane drag is not reliable, as discussed above.

\begin{figure}[t]
\centering
\includegraphics[width=\textwidth]{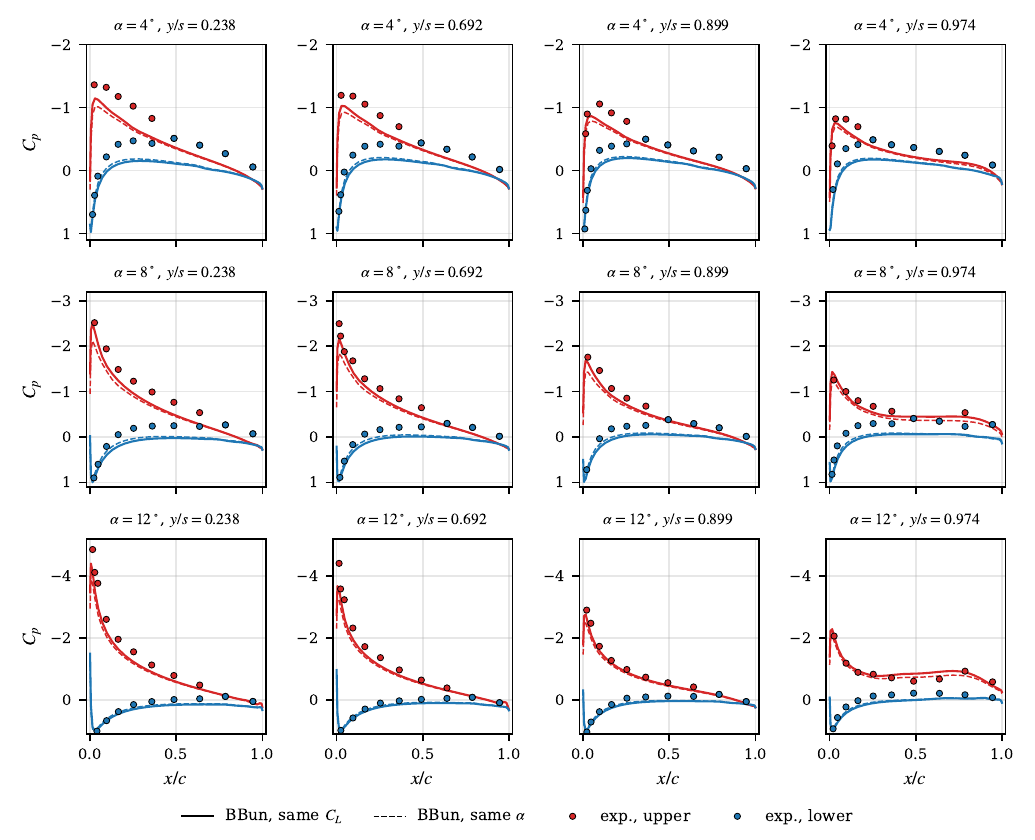}
\caption{NACA~0015 wing, square tip: chordwise pressure at four of the 12
stations of Ref.~\cite{mcalister1991} ($Re=1.5\times10^6$). Symbols:
experiment (upper surface red, lower blue); lines: present method, viscous,
with tip separation, at the incidence that gives the measured wing lift
(solid) and at the same incidence (dashed).}
\label{fig:mcacp}
\end{figure}

\paragraph{Chordwise pressure}
The chordwise pressure distributions at the 12 stations (Fig.~21 of
Ref.~\cite{mcalister1991}, square tip, $Re=1.5\times10^6$, $\alpha=4^\circ$,
$8^\circ$, $12^\circ$) were digitized automatically from the scanned figure:
the frames of the plots are fitted on the (slightly skewed) scan, and the
filled markers of the measurements are isolated by a morphological opening
(accuracy about $\pm0.03$ in $C_p$, checked against the axis ticks).
Figure~\ref{fig:mcacp} compares them with the viscous computations with tip
separation, at the same incidence and at the incidence that gives the
measured wing lift (4.6, 9.1 and $12.9^\circ$), which removes the effect of
the lift difference discussed above. At the same lift the shapes agree along
the whole span: the suction peak, the pressure recovery and, at
$y/s=0.974$--0.994, the suction plateau on the rear half of the upper
surface induced by the tip vortex at $8^\circ$ and $12^\circ$. The measured
pressures are shifted almost uniformly towards negative values on both
surfaces, by 0.23, 0.21 and 0.12 on average at $4^\circ$, $8^\circ$ and
$12^\circ$ ($x/c>0.1$, inboard stations); after removing this shift, which
does not affect the section lift, the rms difference is 0.06--0.08 inboard
and 0.08--0.18 at the three tip stations. The shift is not explained by the
model; the data are not corrected for the walls, and the computed pressure
contains no displacement effect of the boundary layer.

\subsection{Profile drag: NACA 0012 at $Re=6\times10^6$}

\begin{figure}[t]
\centering
\begin{subfigure}[b]{0.42\textwidth}
\includegraphics[width=\textwidth]{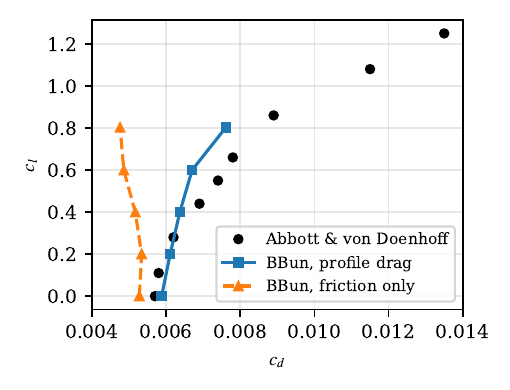}
\caption{}
\label{fig:polar}
\end{subfigure}\hfill
\begin{subfigure}[b]{0.56\textwidth}
\includegraphics[width=\textwidth]{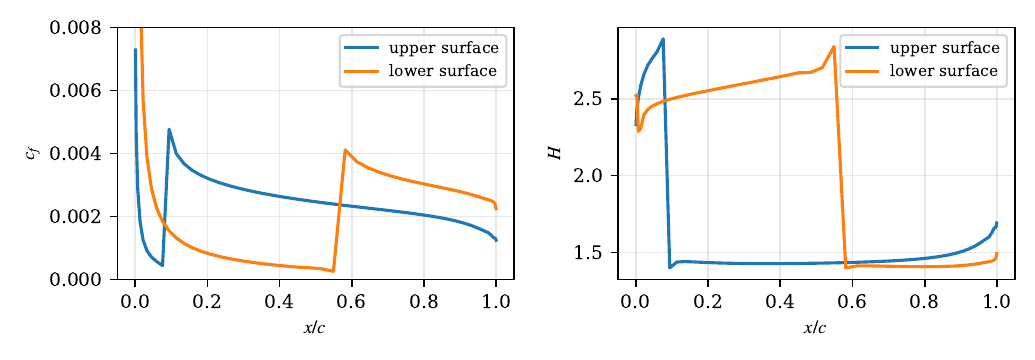}
\caption{}
\label{fig:bl}
\end{subfigure}
\caption{NACA~0012, $Re=6\times10^6$. (a) Profile drag polar. (b) Skin
friction and shape factor at mid span, $\alpha=4^\circ$.}
\end{figure}

The boundary layer is validated on a nearly two-dimensional configuration, a
NACA~0012 wing of aspect ratio 40 (48 panels on each side of the section),
at $Re=6\times10^6$ with free transition, against the data of Abbott and von
Doenhoff for the smooth section~\cite{abbott1959}, digitized from
Ref.~\cite{mason2017}. Figure~\ref{fig:polar} and Table~\ref{tab:polar} show
the profile drag: at zero lift BBun gives $c_d=0.00588$ against 0.0057
($+3\%$), and the agreement is within 5\% up to $c_l=0.4$; at $c_l=0.6$--0.8
the drag rise is underpredicted by about 12\%, which is typical of
Thwaites--Head strip methods. Figure~\ref{fig:bl} shows the skin friction and
the shape factor at $\alpha=4^\circ$: on the upper surface the laminar layer
separates just behind the suction peak and transition occurs at
$x/c\approx0.09$, on the lower surface at $x/c\approx0.56$; at zero lift the
predicted transition is at $x/c\approx0.29$.

\begin{table}[t]
\centering
\caption{Profile drag of the NACA~0012 section, $Re=6\times10^6$.}
\label{tab:polar}
\begin{tabular}{cccc}
\toprule
$c_l$ & BBun & experiment & difference\\
\midrule
0.0 & 0.00588 & 0.0057 & $+3\%$\\
0.2 & 0.00611 & 0.0060 & $+2\%$\\
0.4 & 0.00637 & 0.0067 & $-5\%$\\
0.6 & 0.00670 & 0.0076 & $-12\%$\\
0.8 & 0.00761 & 0.0086 & $-12\%$\\
\bottomrule
\end{tabular}
\end{table}

\subsection{Interacting bodies and box wings}
\label{sec:boxwing}

The interaction between bodies is verified against an independent
vortex-lattice method for several flat rectangular wings (horseshoe
vortices), on rectangular NACA~0006 wings of chord 1 and span 6 at
$\alpha=4^\circ$ (Table~\ref{tab:multi}). Two wings 1000 spans apart
reproduce the isolated wing to all the printed digits. Side by side, and as
the lower wing of a biplane or the front wing of a tandem, the lift ratios
agree within 0.5\%. The upper wing of the biplane ($-3.5\%$) and the rear
wing of the tandem ($-7.4\%$) get less lift than in the vortex-lattice
method, because the present method sheds the wakes along $\V_\infty$ while the
vortex-lattice method keeps them in the plane of each wing: at the rear wing
of the tandem the front wake passes 0.29 chords below it instead of 0.5, and
with this offset the vortex-lattice method gives 0.605 (present 0.604).

\begin{table}[t]
\centering
\caption{Lift of each wing divided by the lift of the isolated wing,
NACA~0006 wings ($c=1$, $b=6$, $\alpha=4^\circ$, inviscid).}
\label{tab:multi}
\begin{tabular}{llcc}
\toprule
layout & wing & present & VLM\\
\midrule
two wings 1000 spans apart & each & 1.000 & 1.000\\
side by side, one chord between the tips & each & 1.040 & 1.038\\
biplane, gap one chord & lower / upper & 0.807 / 0.776 & 0.804 / 0.804\\
tandem, 3 chords behind, 0.5 above & front / rear & 1.033 / 0.604 & 1.038 / 0.652\\
\bottomrule
\end{tabular}
\end{table}

\begin{figure}[t]
\centering
\includegraphics[width=\textwidth]{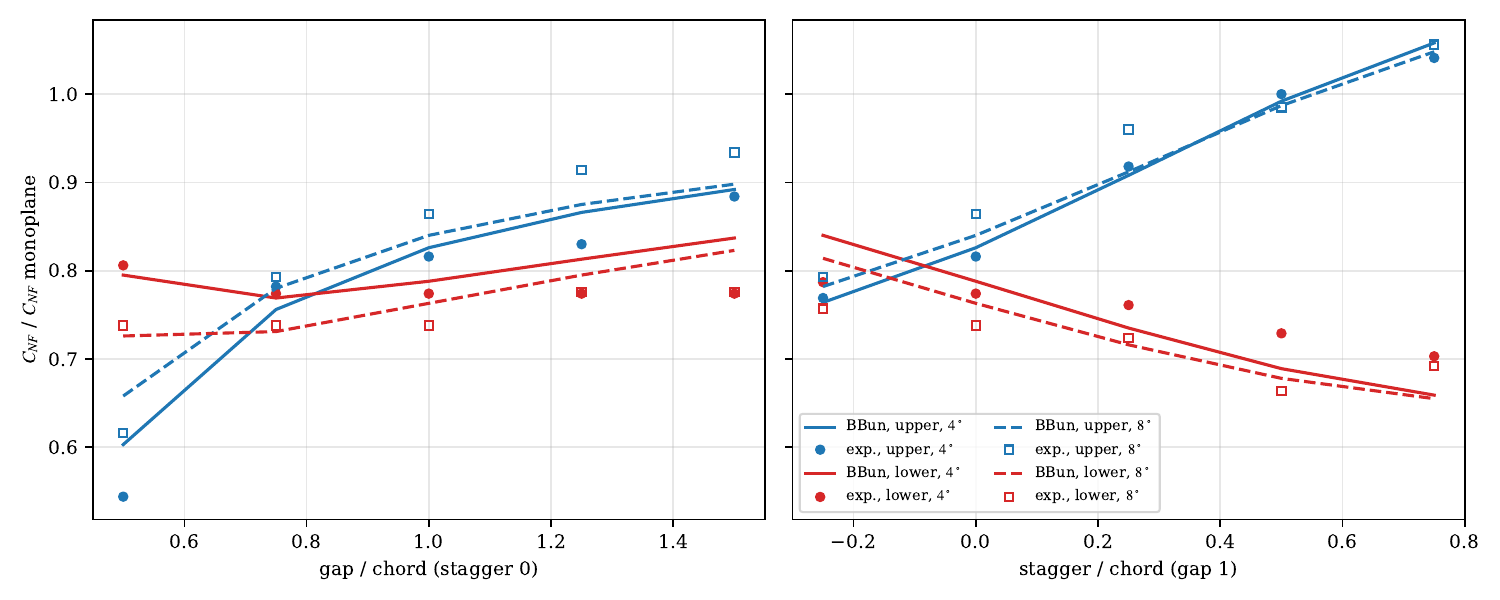}
\caption{Biplane of NACA TN~310~\cite{knight1929} (Clark~Y, aspect ratio 6):
normal-force coefficient of each wing divided by that of the monoplane,
against the gap (left) and the stagger (right), at $4^\circ$ and $8^\circ$.
Lines: present method (inviscid); symbols: experiment ($Re=1.5\times10^5$).}
\label{fig:biplane}
\end{figure}

\begin{figure}[t]
\centering
\includegraphics[width=0.49\textwidth,trim=40 80 40 120,clip]{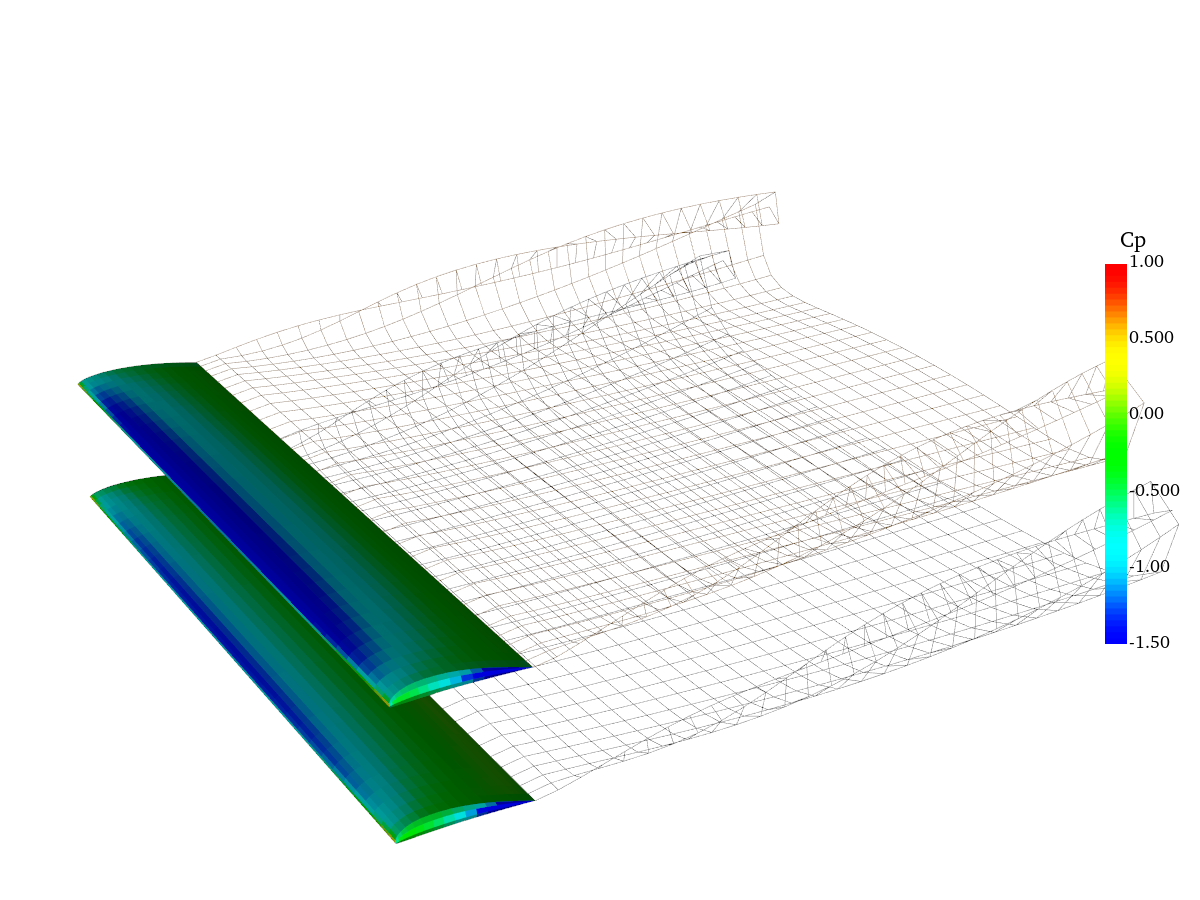}\hfill
\includegraphics[width=0.49\textwidth,trim=40 80 40 120,clip]{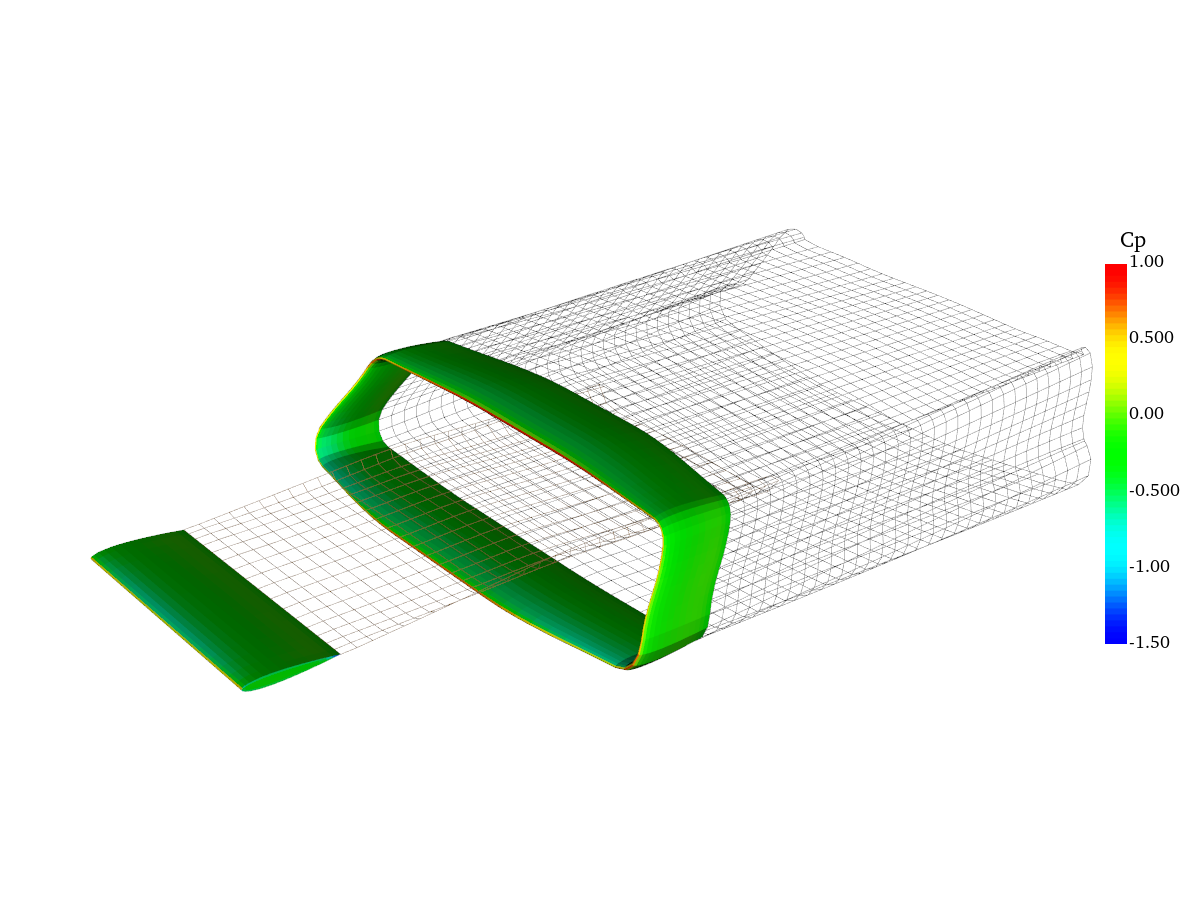}
\caption{Configurations of two interacting bodies: pressure coefficient on the
bodies and free wakes (grey: first body, brown: second body). Left: biplane
of Fig.~\ref{fig:biplane}, gap 1 chord, no stagger, $\alpha=8^\circ$. Right: box
wing with a canard wing ahead, $\alpha=3^\circ$; the canard wake passes
through the box.}
\label{fig:multirender}
\end{figure}

The interaction is validated against the biplane measurements of Knight
and Noyes~\cite{knight1929}: two Clark~Y wings of aspect ratio 6 with round
planform tips, tested as half-span models on a reflection plane, at
$Re=1.5\times10^5$, with the gap varied from 0.5 to 1.5 chords (no stagger)
and the stagger from $-25\%$ to $+75\%$ of the chord (gap 1 chord). The
normal force of each wing was obtained by integrating the pressures (accuracy
$\pm2\%$), without wall corrections. The normal-force coefficients of the two
wings and of the monoplane were digitized from the figures of the report
(reading accuracy about $\pm0.02$). The computations use rectangular wings of
the same planform area (span 5.565 chords, round tips; 2880 panels for the
biplane), inviscid. Since the report compares each wing with the monoplane of
the same model, the comparison is made on the interference ratio
$C_{NF}/C_{NF,\,\mathrm{mono}}$, which also removes most of the Reynolds-number
and wall effects; the monoplane itself agrees in absolute terms
($C_{NF}=0.446$, 0.735, 1.018 at $0^\circ$, $4^\circ$, $8^\circ$, against
0.43--0.45, 0.735--0.775, 0.99--1.07 for the two models).
Figure~\ref{fig:biplane} shows the ratios at $4^\circ$ and $8^\circ$. The
computation reproduces the strong reduction of the upper-wing load at small
gaps, the recovery with the gap, and the transfer of load from the lower to
the upper wing with positive stagger, with the upper wing exceeding the
monoplane for large stagger. At $4^\circ$ and $8^\circ$ the ratios agree
within 0.03 (upper wing) and 0.04 (lower wing) rms, comparable with the
reading accuracy, and the share of the load carried by the upper wing within
0.02 rms. At $0^\circ$ the computed ratios are higher (by 0.08--0.09 rms),
where the lift of the Clark~Y at this low Reynolds number is most sensitive
to the boundary layer. Figure~\ref{fig:multirender} shows the biplane with
its wakes and a box wing with a canard in front of it, whose wake passes
through the box.

For ring-shaped bodies the interior velocity conditions of
Section~\ref{sec:multi} remove the spurious link flow: on the random box
wings where it had appeared, the lowest $C_p$ went from $-58$, $-205$, $-47$,
$-24$ and $-10.5$ to values between $-1.0$ and $-1.3$. On box wings without it
the lift changes by 0.2--0.8\%, and on the default box wing of the grid
generator ($\alpha=5.7^\circ$) the difference from the plain Neumann solution
decreases with the grid: 0.22\%, 0.11\% and 0.07\% on 2560, 5760 and 10240
panels, while both solutions converge in the same way (+3.0\% and +1.5\% at
the two refinements). In random campaigns of 80 box wings with the parameter
ranges of a design study (chords 0.4--1.5, link chord 0.3--1 times the
smaller wing chord, spans 3--12, link span 0.5--3, viscous, 2560 panels) the
spurious link flow had made 16 points unusable; with the interior conditions
none, and the largest surface speed of all the points was
$2.03|\V_\infty|$.

\paragraph{Symmetry and a wing--fuselage combination}
Whole configurations and their halves computed with the image method
(Section~\ref{sec:sym}) give the same loads to all the printed digits (eight)
for a rectangular wing with and without tip separation, the default box
wing and two counter-rotating propellers, and the objective function of the
box-wing design study gives the same efficiency on random design points in
0.56 of the time. With the boundary layer the profile drag of a wing differs
by $10^{-3}$, which is the left--right asymmetry of the whole wing itself
(the stagnation panel of a chordwise strip is assigned to one side, and the
strips of the two halves run in opposite directions). A semispan wing of
aspect ratio 8 beside a fuselage of radius 0.4 chords (root 0.05 chords from
its side, $\alpha=4^\circ$, inviscid) gives $C_L=0.309$ on the wing and 0.019
on the non-lifting fuselage, 0.327 in total, against 0.320 for the continuous
wing and 0.270 for the same wing without the fuselage, whose gap at the root
produces a root vortex: the classical result that the lift of a wing--body
combination is close to that of the continuous wing.

\begin{figure}[t]
\centering
\includegraphics[width=\textwidth]{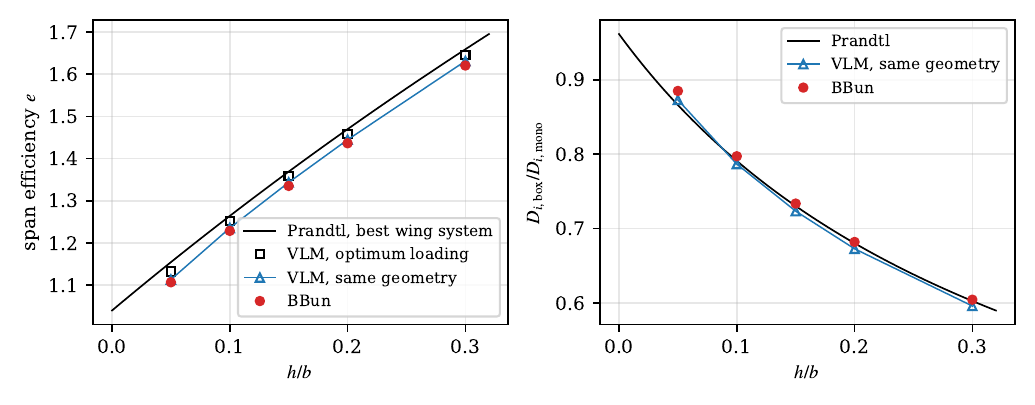}
\caption{Box wings of span 8 chords, equal wings, vertical links,
$\alpha=4^\circ$: span efficiency (left) and induced drag relative to the
monoplane of the same span and lift (right) against the height-to-span ratio.
Present method (far-field lift and drag), vortex-lattice method of the same
geometry and its optimum loading, and Prandtl's best wing system.}
\label{fig:prandtl}
\end{figure}

\paragraph{Induced drag of box wings}
For the ratio of the induced drag of the optimally loaded box wing (the
``best wing system'') to that of the elliptically loaded monoplane of the
same span and lift, Prandtl~\cite{prandtl1924} gave the approximation
$D_{i,\mathrm{box}}/D_{i,\mathrm{mono}}=(1+0.45\,h/b)/(1.04+2.81\,h/b)$, with $h$
the height of the box (see also Ref.~\cite{frediani2009}).
Box wings of span $b=8$ chords with equal wings, vertical links, NACA~0006
sections and corner radius 0.15 chords were computed for
$h/b=0.05$--0.3 at $\alpha=4^\circ$ (inviscid, about 4100--5000 panels), with
a rectangular monoplane of the same span and section. The rectangular wings
are not optimally loaded, so the comparison is made in two steps
(Fig.~\ref{fig:prandtl}). An independent vortex-lattice method on the same
path in the $y$--$z$ plane (flat mean surfaces, horseshoe vortices) gives the
span efficiency of the same geometry and, by minimizing the Trefftz-plane
drag of the strip circulations for the same lift, the optimum of the path:
the latter is 0.7--1.9\% below Prandtl's approximation (whose accuracy is of
this order; the corners are also rounded here). The present method, with
lift and drag both from the far field, gives $e$ 0.5--0.7\% below the
vortex-lattice method for the box wings and 0.8\% above for the monoplane
($e=0.980$ against 0.972), and changes by $-0.08\%$ when the grid is refined
(panel width 0.1 to 0.07 chords, 12 to 16 panels per side of the section).
The resulting drag ratio, 0.885, 0.797, 0.734, 0.682 and 0.605 for
$h/b=0.05$, 0.1, 0.15, 0.2 and 0.3, is within 0.4\% of Prandtl's ratio for
$h/b\ge0.1$ and 2\% at $h/b=0.05$. According to the vortex-lattice method,
the rectangular box wings are 0.9--1.7\% below the optimum of their path and
the rectangular monoplane 2.8\% below the elliptic loading; in the ratio these
differences and the approximation of the formula largely compensate.

\subsection{Accuracy of the pressure loads}
\label{sec:accuracy}

This subsection quantifies the main limitation of the discretization
(Table~\ref{tab:limits}). The pressure lift converges to the far-field lift $\rho|\V_\infty|\oint\Gamma\,\mathrm dy$
at first order in the chordwise panel size, from below, while the far-field
lift changes little. On a quasi-two-dimensional NACA~0012 wing (aspect ratio
20) the far-field section lift changes by only 0.7\% from 16 to 64 panels per
side, while the error of the pressure lift, $-8.1$, $-5.5$, $-4.1$, $-2.8$ and
$-2.1\%$ with 16, 24, 32, 48 and 64 panels per side, halves with the panel
size, about $-130\%/n$ with $n$ panels per side. The monoplane of span 8 gives
$-5.9\%$ and $-4.5\%$ with 24 and 32 panels per side, and the box wing with
$h/b=0.1$ $-12.3\%$ and $-9.3\%$ with 12 and 16: the same behaviour, slightly
larger on the box wing. The span efficiency depends only on the shape of the
circulation and converges quickly, provided that lift and drag are taken
consistently from the far field; the pressure lift is appropriate for the
loads, but on coarse chordwise grids it is several per cent low. On the
quasi-two-dimensional wing the integration of the pressure is exact to
0.03\%, and the error is a nearly uniform deficit of the surface speed on
both surfaces (1.6\% of $|\V_\infty|$ with 32 panels per side, against an
independent two-dimensional panel method at the same lift), with a
corresponding flow inside the body that also vanishes at first order. It is
therefore a first-order error of the near-body velocity field of the
constant-strength vortex rings with the Neumann condition, not of the
evaluation of the surface velocity (extrapolating the velocity from outside
the surface gives the same deficit) nor of the circulation; interior velocity
conditions on ordinary wings do not remove it. A Dirichlet (internal
potential) formulation would be the remedy. With thin sections (NACA~0006) the pressure also shows a
suction peak at the trailing edge on both surfaces ($C_p\approx-1$ over the
last 3\% of the chord, absent for $t/c\ge0.12$), which affects the
near-field drag but not the lift.

\subsection{Unsteady pressure and rotating bodies}

\begin{figure}[t]
\centering
\includegraphics[width=0.7\textwidth]{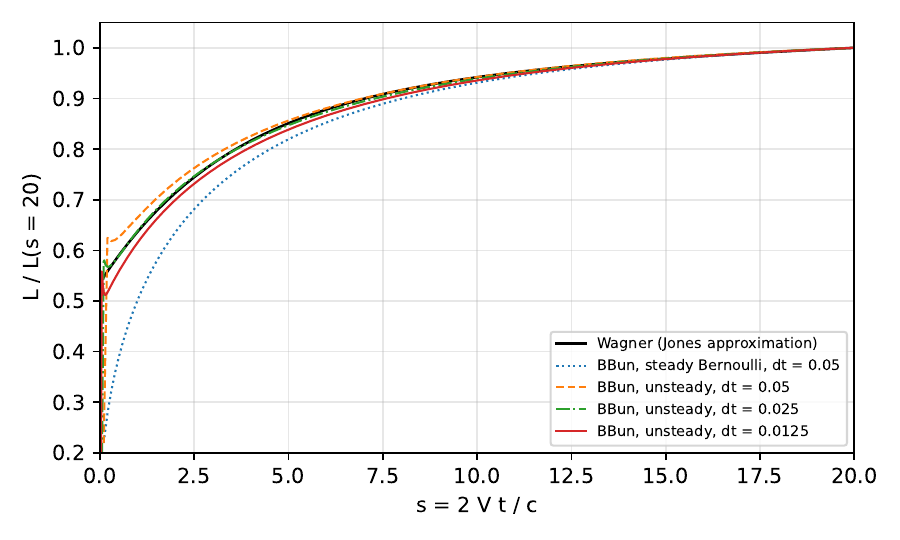}
\caption{Impulsively started wing of aspect ratio 40: pressure lift divided
by its value at $s=20$, with the steady and the unsteady Bernoulli equation,
against Wagner's function.}
\label{fig:wagner}
\end{figure}

\paragraph{Impulsive start}
A quasi-two-dimensional wing (aspect ratio 40, NACA~0012, 32 panels per
side, $\alpha=4^\circ$) started impulsively is compared with Wagner's function
(Fig.~\ref{fig:wagner}). With the steady Bernoulli equation the pressure
lift starts from 0.28 of its value at $s=2Vt/c=20$, against 0.56 of the
Wagner function, and is 5\% low at $s=4$. With the unsteady term,
Eq.~(\ref{eq:cpunst}), the history follows Wagner's function: at $s=4$ it
gives 0.826, 0.815 and 0.804 with $V\Delta t/c=0.05$, 0.025 and 0.0125,
against 0.816, and within 1.5\% of it from $s=1$ on with all three steps; at
$s=0.2$ it gives 0.625, 0.568 and 0.517 against 0.559. With the smallest step
the response is slightly slower than Wagner's, as expected for a finite
aspect ratio. At $s=20$ the unsteady term changes the lift by 0.5\%.

\paragraph{Verification of the motion}
A wing moving on a circle of radius $10^5$ chords reproduces the fixed wing
in the equivalent free stream to $1.1\times10^{-4}$ over the whole transient,
and to $1.1\times10^{-5}$ with a radius of $10^6$ chords, i.e.\ to the order
of $c/R$; the rigid path and the path with the influence matrix rebuilt at
every step agree to $1.6\times10^{-8}$ on a propeller. The lift from the
circulation of the moving wing agrees with the far-field lift of the fixed
wing to $4\times10^{-4}$.

\begin{figure}[t]
\centering
\includegraphics[width=0.85\textwidth,trim=250 100 150 150,clip]{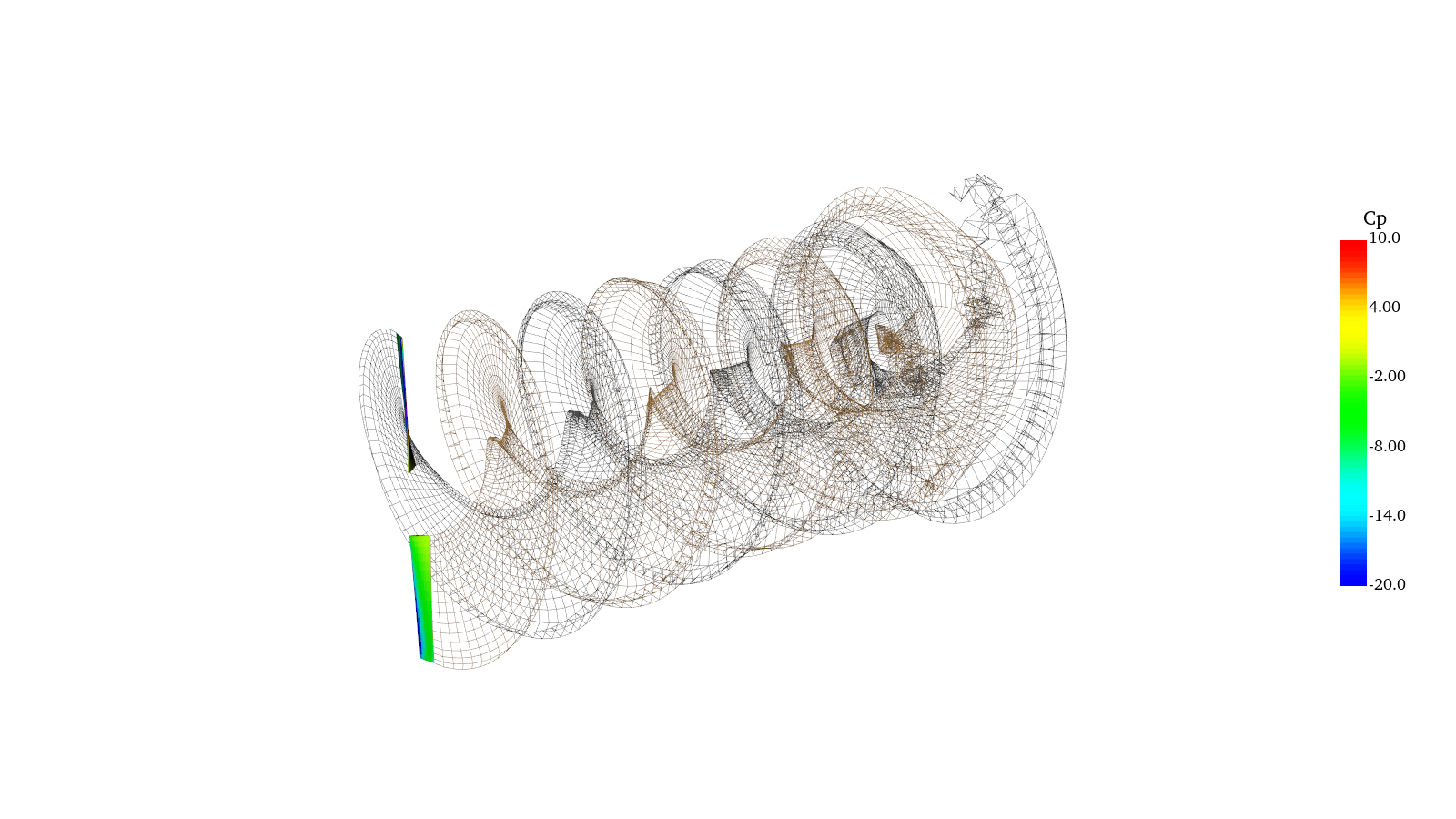}
\caption{Two-bladed propeller after four revolutions ($J=0.5$): pressure
coefficient on the blades and the free wake sheets, with the roll-up of the
tip vortices and the root vortex.}
\label{fig:prop}
\end{figure}

\paragraph{A propeller}
A two-bladed propeller (radius $R$, blades from $0.2R$ to $R$ with chord
$0.12R$, NACA~0012, constant geometric pitch $P/D=0.8$) is computed at the
advance ratio $J=V/(nD)=0.5$, inviscid, with $5^\circ$ of azimuth per step;
the loads converge within 0.2\% in three revolutions (Fig.~\ref{fig:prop}).
Table~\ref{tab:prop} compares the loads from the pressure and from the
circulation with blade-element momentum theory (Prandtl tip loss, section
lift slope $2\pi(1+0.77\,t/c)$, no drag). The pressure loads have the
first-order error of Section~\ref{sec:boxwing}, larger on the torque (an
in-plane, drag-like component): the efficiency is 0.63, 0.77 and 0.82 with
12, 24 and 48 panels per side, 0.86 extrapolated. The loads from the
circulation change by 1.3\% (thrust) between 12 and 48 panels per side, with
the efficiency 0.857--0.856, the extrapolated value of the pressure loads.
Blade-element momentum theory gives a thrust 5\% higher, as is usual for this
theory, and the efficiency 0.871.

\begin{table}[t]
\centering
\caption{Two-bladed propeller, $P/D=0.8$, $J=0.5$, inviscid: thrust and
torque coefficients $C_T=T/(\rho n^2D^4)$, $C_P=P/(\rho n^3D^5)$ and efficiency.}
\label{tab:prop}
\begin{tabular}{l ccc ccc}
\toprule
& \multicolumn{3}{c}{from the pressure} & \multicolumn{3}{c}{from the circulation}\\
\cmidrule(lr){2-4}\cmidrule(lr){5-7}
panels per side & $C_T$ & $C_P$ & $\eta$ & $C_T$ & $C_P$ & $\eta$\\
\midrule
12 & 0.0470 & 0.0372 & 0.632 & 0.0542 & 0.0316 & 0.857\\
24 & 0.0510 & 0.0333 & 0.765 & 0.0548 & 0.0320 & 0.856\\
48 & 0.0528 & 0.0322 & 0.819 & 0.0550 & 0.0321 & 0.856\\
\midrule
BEMT & \multicolumn{6}{c}{$C_T=0.0577$, $C_P=0.0331$, $\eta=0.871$}\\
\bottomrule
\end{tabular}
\end{table}

\subsection{A complete aircraft}
\label{sec:aircraft}

\begin{figure}[t]
\centering
\includegraphics[width=\textwidth,trim=180 60 40 100,clip]{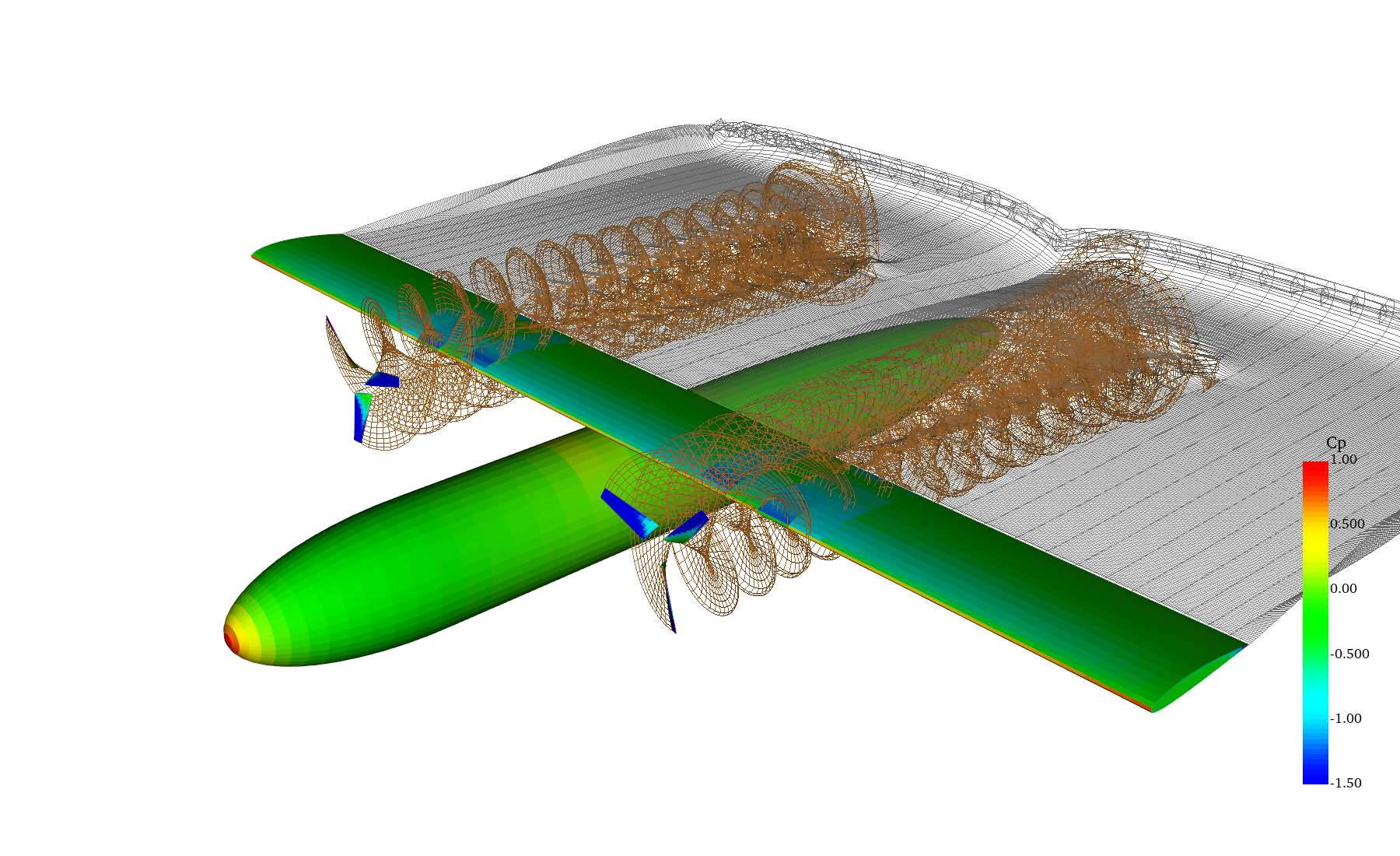}
\caption{Light twin-engine aircraft with counter-rotating tractor
propellers, $\alpha=3^\circ$, $J=0.8$, after five time units (250 steps):
pressure coefficient on the bodies and free wakes (grey: wing, brown:
propellers). Half model with the image method, shown whole.}
\label{fig:aircraft}
\end{figure}

As an example of the whole method, Fig.~\ref{fig:aircraft} shows a light
twin-engine aircraft in steady flight: a fuselage, a high wing and two
counter-rotating tractor propellers in front of it. The fuselage is a body
of revolution of length 8 chords and radius 0.5 chords (ogive nose, tail
cone), without wake (Section~\ref{sec:sym}); the wing, NACA~2412 with aspect
ratio 10, lies 0.06 chords above the fuselage, continuous across the plane of
symmetry, so that it has no root vortex (a wing ending at the side of the
fuselage would shed a strong trailing vortex along it). Each propeller has
three blades of radius 0.6 chords (pitch-to-diameter ratio 1.0), its hub 0.8
chords ahead of the wing leading edge at 1.8 chords from the plane of
symmetry, and runs at the advance ratio $J=0.8$, with the outboard blade
moving up. The flight speed is $|\V_\infty|=1$ at $\alpha=3^\circ$, inviscid. Only the
half $y\ge0$ is computed (2728 panels): its mirror image supplies the other
half of the fuselage and of the wing and the second propeller, rotating the
other way. Since the propellers move relative to the rest, the influence
matrix is rebuilt at every step; the pressure includes the unsteady term,
Eq.~(\ref{eq:cpunst}), for the wing and the fuselage in the unsteady
slipstreams, and the collision treatment of Section~\ref{sec:coll} keeps the
slipstreams out of the wing (after 100 steps 121 of their nodes lay inside
the wing without it, none with it, at a negligible cost). The run covers
five time units, 250 steps of $7.5^\circ$ of propeller azimuth, so that the
slipstreams reach 3.7 chords behind the wing trailing edge; it took four
minutes on 16 cores.

The figure shows the stagnation point on the nose of the fuselage, the
suction on the wing, stronger behind the propellers, the helical sheets of
the blades, with their tip vortices rolling up, which pass around the wing
and continue far behind it, and the wing wake deformed by the slipstreams.
At the end of the run the configuration has $C_L=0.100$ referred to the
wetted area and dynamic pressure, and each propeller a mean thrust (from the
circulation) of $0.148\,\rho V^2c^2$.

\subsection{Computational performance}

\begin{figure}[t]
\centering
\includegraphics[width=0.5\textwidth]{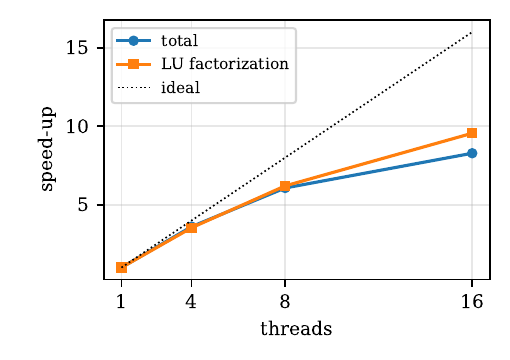}
\caption{Parallel speed-up, 9792 panels.}
\label{fig:speedup}
\end{figure}

Table~\ref{tab:perf} and Fig.~\ref{fig:speedup} report the wall-clock time
of a rectangular wing with 9792 panels for 40 time steps on a 16-core
processor (AMD Ryzen 9 9950X, gfortran~15, OpenBLAS~0.3). The bordered LU
factorization takes 1.28\,s on 16 threads, against 312\,s for a singular
value decomposition of the same matrix, which parallelizes poorly; the
evaluation of the wake as a list of vortex lines halves the cost of the
Biot--Savart sums. On one thread the whole computation is 4.4 times faster
than a variant with the singular value decomposition and the sum over rings;
on 16 threads it is 36 times faster than that variant on one thread, and the
parallel speed-up of the present method is 8.3 (the timings vary by about
3\% between repeated runs). With relative motion between the bodies the
influence matrix is rebuilt and factorized at every step: for the aircraft of
Section~\ref{sec:aircraft} (2728 panels) this takes 0.2\,s per step, and the
velocity of the wake nodes, which grows with the wake, dominates the cost.

\begin{table}[t]
\centering
\caption{Wall-clock time [s], 9792 panels, 40 time steps.}
\label{tab:perf}
\begin{tabular}{lcccc}
\toprule
threads & 1 & 4 & 8 & 16\\
\midrule
present method, total & 96.2 & 26.7 & 15.9 & 11.6\\
present method, LU factorization & 12.2 & 3.4 & 2.0 & 1.3\\
SVD and sum over rings, total & 424.1 & -- & -- & 321.7\\
\bottomrule
\end{tabular}
\end{table}

\section{Summary and recommendations for use}
\label{sec:concl}

An unsteady vortex-ring panel method with free wake, implicit unsteady Kutta
condition and strip-wise viscous correction has been described and
validated, as a reference for the code BBun. The formulation is
classical; the attention has been devoted to the consistency of its
discretization. The main findings are:
\begin{itemize}
\item the implicit Kutta condition with a fixed shedding panel, and the
  exclusion of any velocity evaluation on the separation line, give grid
  convergence of the loads, which an explicit condition does not
  (\ref{app:earlier});
\item the wake kernel must be applied line by line: evaluating the two rings
  of a shared line with different kernels leaves a spurious vortex of the
  order of the circulation, which made the steady lift depend on the length
  of the first wake row by a factor 2.5;
\item the singular Neumann problem of a closed body is solved exactly (as the
  minimum-norm solution) by a bordered LU factorization with the projection
  of the right-hand side on the range of the matrix, at a small fraction of
  the cost of a singular value decomposition;
\item the body rings must be desingularized when the wake nodes are
  convected: with the exact kernel the tip sheets, which run along the tip
  caps, get spurious velocities and irregular loads;
\item several bodies are handled in one system with one gauge condition per
  closed body; on ring-shaped bodies (box wings) interior velocity
  conditions, solved in the least-squares sense with a small weight, remove
  a spurious flow along thin links and do not change the converged solution;
  the interference of two wings is validated against biplane measurements:
  the load of each wing relative to the monoplane agrees within 0.03--0.04
  rms over gaps of 0.5--1.5 chords and staggers of $-25$ to $+75\%$;
\item the induced drag of box wings relative to the monoplane agrees within
  0.4\% with Prandtl's best wing system for $h/b\ge0.1$ (2\% at $h/b=0.05$),
  and their span efficiency within 0.7\% with an independent vortex-lattice
  method, when lift and drag are both taken from the far field; the pressure
  lift converges to the far-field lift at first order in the chordwise panel
  size and is several per cent low on coarse grids;
\item the section lift agrees within 1\% with inviscid theory in the
  two-dimensional limit, and the three-dimensional lift slope agrees with an
  independent vortex-lattice method; on the McAlister--Takahashi wing the lift
  is lower than the measured one by an amount consistent with the uncorrected
  wall interference of the experiment, and at the same lift the chordwise
  pressure distributions, including the effect of the tip vortex, agree
  apart from a nearly uniform shift of the measured pressures;
\item the induced drag must be computed in the Trefftz plane: induced plus
  form drag agree with the measured pressure drag within about 10\%, while
  the near-field pressure integration overestimates the drag;
\item the integral boundary layer predicts the profile drag of the NACA~0012
  section within 3\% at zero lift and 12\% at $c_l\approx0.7$;
\item with the unsteady term of the Bernoulli equation, formulated for the
  potential of the doublet method and for moving bodies, the lift of an
  impulsively started wing converges to Wagner's function; propeller loads
  computed from the circulation converge with the chordwise panels much
  faster than the pressure loads, and give the efficiency of blade-element
  momentum theory within 2\%; symmetric configurations are computed on
  their half by the image method with identical results.
\end{itemize}
The separation lines are derived from the grid connectivity, which makes the
method directly applicable to closed lifting systems such as the box wing,
whose wake is a closed band without tips, and to configurations of several
interacting bodies. Tip separation from square tips improves the loading near
the tip, but its Trefftz-plane drag is not reliable; its flow is regular only
if the body rings are desingularized when the wake nodes are convected.

\paragraph{Recommendations for use}
The verification and validation of Section~\ref{sec:results} lead to the
following rules for the applications of the code.
\begin{itemize}
\item Lift-to-drag ratios and span efficiencies: take the lift from the
  circulation (far-field, Kutta--Joukowski) and the drag from the Trefftz
  plane plus the profile drag of the boundary layer; do not combine the
  pressure lift with the Trefftz-plane drag.
\item Pressure loads and distributions: use at least 32 panels on each side
  of the sections (a pressure lift about 4\% low) and more for accurate
  values; the chordwise shapes are reliable on coarser grids.
\item Propellers: use the loads from the circulation, which converge with
  12 panels per side; the pressure loads, the torque above all, need 48 or
  more. Use an azimuth step of 3--$7.5^\circ$.
\item Unsteady flows (starts, propeller slipstreams, oblique inflow): switch
  on the unsteady term of the Bernoulli equation; in steady conditions it
  vanishes.
\item Wakes reaching bodies (a slipstream on a wing): switch on the
  collision treatment; it is cheap when the wakes stay away from most of the
  bodies, and unnecessary otherwise.
\item Symmetric configurations: compute the half with the image method;
  the results are those of the whole configuration.
\item Wings and fuselages: since bodies are not intersected, prefer
  configurations in which the wing is continuous (high or low wing across
  the plane of symmetry) to a wing ending at the side of the fuselage, whose
  root vortex runs along the fuselage.
\item Box wings: keep the interior velocity conditions (default); check the
  largest surface speed of the solution (a spurious flow shows speeds well
  above $2.5|\V_\infty|$).
\end{itemize}

\paragraph{Outlook}
A Dirichlet (internal potential) formulation would remove the first-order
error of the surface velocity and the interior flow of the Neumann solution;
other extensions are a three-dimensional or wake-including viscous model, a
deformation of the wakes around the bodies, and the intersection of bodies
(wing--fuselage junctions).

\section*{Acknowledgements}
This work is dedicated to the memory of Maurizio Landrini, who left us
suddenly in 2003. We wrote the first version of this code together, while I
was working on my thesis, and what is described in these pages still grows
from that time spent working side by side. It all began there.

\appendix
\section{Lessons from an earlier implementation}
\label{app:earlier}

The present code is a rewrite of an implementation of the same method written
some twenty years earlier in Fortran~77. In that implementation the lift did
not converge under grid refinement: it increased by 16\% and then by 7\% at
each refinement of a wing at high incidence, with no sign of a limit
(Fig.~\ref{fig:conv}(a)). The analysis identified three causes:
\begin{itemize}
\item the unsteady Kutta condition was applied explicitly, with the
  circulation of the preceding time step, so that the trailing-edge vortex
  was unbalanced by an amount whose effect on the neighbouring collocation
  points grows as the panels shrink;
\item the first wake row was placed with the velocity computed on the
  trailing edge itself, where it is singular;
\item the wake-to-body influence was evaluated with the singular kernel.
\end{itemize}
While correcting them, a further source of parameter dependence was found in
the discretization of the wake itself: two rings of a shared wake line
evaluated with different kernels (Section~\ref{sec:consistency}), which made
the steady lift depend on the length of the first wake row by a factor 2.5
(Table~\ref{tab:sens}). The null space of the matrix of a closed body was
handled with the pseudo-inverse from a singular value decomposition, whose
cost dominated the computation (Table~\ref{tab:perf}); the test case
distributed with that code (1012 panels, 50 steps) runs in 2.2\,s with the
present one instead of 162\,s. The collocation point was the point of
minimum self-induced velocity of the panel, which with the settings used
there always coincided with the centroid; the present code uses the
centroid.

\bibliographystyle{elsarticle-num}
\bibliography{refs}

\begin{thebibliography}{10}
\expandafter\ifx\csname url\endcsname\relax
  \def\url#1{\texttt{#1}}\fi
\expandafter\ifx\csname urlprefix\endcsname\relax\def\urlprefix{URL }\fi
\expandafter\ifx\csname href\endcsname\relax
  \def\href#1#2{#2} \def\path#1{#1}\fi

\bibitem{hess1967}
J.~L. Hess, A.~M.~O. Smith, Calculation of potential flow about arbitrary
  bodies, Progress in Aerospace Sciences 8 (1967) 1--138.

\bibitem{morino1974}
L.~Morino, C.-C. Kuo, Subsonic potential aerodynamics for complex
  configurations: a general theory, AIAA Journal 12~(2) (1974) 191--197.

\bibitem{maskew1987}
B.~Maskew, Program {VSAERO} theory document, Tech. Rep. CR-4023, NASA (1987).

\bibitem{ashby1991}
D.~L. Ashby, M.~R. Dudley, S.~K. Iguchi, L.~Browne, J.~Katz, Potential flow
  theory and operation guide for the panel code {PMARC}, Tech. Rep. TM-102851,
  NASA (1991).

\bibitem{magnus1980}
A.~E. Magnus, M.~A. Epton, {PAN AIR} -- a computer program for predicting
  subsonic or supersonic linear potential flows about arbitrary configurations
  using a higher order panel method. {Volume I}: Theory document, Tech. Rep.
  CR-3251, NASA (1980).

\bibitem{hess1990}
J.~L. Hess, Panel methods in computational fluid dynamics, Annual Review of
  Fluid Mechanics 22 (1990) 255--274.

\bibitem{erickson1990}
L.~L. Erickson, Panel methods -- an introduction, Tech. Rep. TP-2995, NASA
  (1990).

\bibitem{katz2001}
J.~Katz, A.~Plotkin, Low-Speed Aerodynamics, 2nd Edition, Cambridge University
  Press, Cambridge, 2001.

\bibitem{tugnoli2021}
M.~Tugnoli, D.~Montagnani, M.~Syal, G.~Droandi, A.~Zanotti, Mid-fidelity
  approach to aerodynamic simulations of unconventional {VTOL} aircraft
  configurations, Aerospace Science and Technology 115 (2021) 106804.

\bibitem{mcdonald2022}
R.~A. McDonald, J.~R. Gloudemans, Open vehicle sketch pad: An open source
  parametric geometry and analysis tool for conceptual aircraft design, in:
  AIAA SciTech 2022 Forum, 2022, aIAA 2022-0004.

\bibitem{winckelmans1993}
G.~S. Winckelmans, A.~Leonard, Contributions to vortex particle methods for the
  computation of three-dimensional incompressible unsteady flows, Journal of
  Computational Physics 109 (1993) 247--273.

\bibitem{alvarez2020}
E.~J. Alvarez, A.~Ning, High-fidelity modeling of multirotor aerodynamic
  interactions for aircraft design, AIAA Journal 58~(10) (2020) 4385--4400.

\bibitem{prandtl1924}
L.~Prandtl, Induced drag of multiplanes, Tech. Rep. TN-182, NACA (1924).

\bibitem{frediani2009}
A.~Frediani, G.~Montanari, Best wing system: an exact solution of the
  {P}randtl's problem, in: Variational Analysis and Aerospace Engineering,
  Vol.~33 of Springer Optimization and Its Applications, Springer, New York,
  2009, pp. 183--211.

\bibitem{vatistas1991}
G.~H. Vatistas, V.~Kozel, W.~C. Mih, A simpler model for concentrated vortices,
  Experiments in Fluids 11 (1991) 73--76.

\bibitem{leishman2002}
J.~G. Leishman, M.~J. Bhagwat, A.~Bagai, Free-vortex filament methods for the
  analysis of helicopter rotor wakes, Journal of Aircraft 39~(5) (2002)
  759--775.

\bibitem{lapack1999}
E.~Anderson, Z.~Bai, C.~Bischof, S.~Blackford, J.~Demmel, J.~Dongarra,
  J.~Du~Croz, A.~Greenbaum, S.~Hammarling, A.~McKenney, D.~Sorensen, {LAPACK}
  Users' Guide, 3rd Edition, SIAM, Philadelphia, 1999.

\bibitem{thwaites1949}
B.~Thwaites, Approximate calculation of the laminar boundary layer,
  Aeronautical Quarterly 1 (1949) 245--280.

\bibitem{cebeci1977}
T.~Cebeci, P.~Bradshaw, Momentum Transfer in Boundary Layers, Hemisphere,
  Washington, 1977.

\bibitem{michel1951}
R.~Michel, {\'E}tude de la transition sur les profils d'aile; {\'e}tablissement
  d'un crit{\`e}re de d{\'e}termination du point de transition et calcul de la
  tra{\^\i}n{\'e}e de profil incompressible, Tech. Rep. 1/1578A, ONERA (1951).

\bibitem{head1958}
M.~R. Head, Entrainment in the turbulent boundary layer, Reports and Memoranda
  3152, Aeronautical Research Council (1958).

\bibitem{ludwieg1949}
H.~Ludwieg, W.~Tillmann, Untersuchungen {\"u}ber die {W}andschubspannung in
  turbulenten {R}eibungsschichten, Ingenieur-Archiv 17 (1949) 288--299.

\bibitem{squire1938}
H.~B. Squire, A.~D. Young, The calculation of the profile drag of aerofoils,
  Reports and Memoranda 1838, Aeronautical Research Council (1938).

\bibitem{lighthill1958}
M.~J. Lighthill, On displacement thickness, Journal of Fluid Mechanics 4~(4)
  (1958) 383--392.

\bibitem{openmp2018}
{OpenMP Architecture Review Board}, {OpenMP} application programming interface,
  version 5.0, \url{https://www.openmp.org} (2018).

\bibitem{openblas2012}
Z.~Xianyi, W.~Qian, Z.~Yunquan, Model-driven level 3 {BLAS} performance
  optimization on {L}oongson 3{A} processor, in: IEEE 18th International
  Conference on Parallel and Distributed Systems (ICPADS), 2012, pp. 684--691.

\bibitem{vtk}
W.~Schroeder, K.~Martin, B.~Lorensen, The visualization toolkit: An
  object-oriented approach to 3d graphics, Kitware, 4th edition (2006).

\bibitem{mcalister1991}
K.~W. McAlister, R.~K. Takahashi, {NACA} 0015 wing pressure and trailing vortex
  measurements, Tech. Rep. TP-3151 (AVSCOM TR 91-A-003), NASA (1991).

\bibitem{nalu}
{Nalu-Wind developers}, Lessons learned from meshing the {McAlister} case, Nalu
  documentation,
  \url{https://nalu.readthedocs.io/en/latest/source/user/nalu_run/McAlisterLessonsLearned.html},
  accessed September 2026.

\bibitem{barlow1999}
J.~B. Barlow, W.~H. Rae, A.~Pope, Low-Speed Wind Tunnel Testing, 3rd Edition,
  Wiley, New York, 1999.

\bibitem{abbott1959}
I.~H. Abbott, A.~E. von Doenhoff, Theory of Wing Sections, Dover, New York,
  1959.

\bibitem{mason2017}
W.~H. Mason, Curiosity number 8: airfoil drag ({NACA} 0012) added to induced
  drag, Virginia Tech, Dept. of Aerospace and Ocean Engineering,
  \url{https://archive.aoe.vt.edu/mason/Mason_f/C80012AirfoilDragaddedtoinduced.pdf}
  (2017).

\bibitem{knight1929}
M.~Knight, R.~W. Noyes, Wind tunnel pressure distribution tests on a series of
  biplane wing models. {Part I}: effects of changes in stagger and gap,
  Technical Note 310, NACA (1929).

\end{thebibliography}
\end{document}